\documentclass[11pt]{article}
\usepackage{epsf}
\def\beq{\begin{eqnarray}}
\def\eeq{\end{eqnarray}}
\def\bea{\begin{eqnarray*}}
\def\eea{\end{eqnarray*}}
\def\NPB#1#2#3{Nucl. Phys. {\bf B#1}, #3 (#2)}
\def\PLB#1#2#3{Phys. Lett. B {\bf #1}, #3 (#2)}

\def\PRD#1#2#3{Phys. Rev. {\bf D#1}, #3 (#2)}

\def\PRL#1#2#3{Phys. Rev. Lett. {\bf #1}, #3 (#2)}
\def\PREP#1#2#3{Phys. Rep. {\bf #1}, #3 (#2)}

\def\centeron#1#2{{\setbox0=\hbox{#1}\setbox1=\hbox{#2}\ifdim
\wd1>\wd0\kern.5\wd1\kern-.5\wd0\fi
\copy0\kern-.5\wd0\kern-.5\wd1\copy1\ifdim\wd0>\wd1
\kern.5\wd0\kern-.5\wd1\fi}}
\def\ltap{\;\centeron{\raise.35ex\hbox{$<$}}{\lower.65ex\hbox{$\sim$}}\;}
\def\gtap{\;\centeron{\raise.35ex\hbox{$>$}}{\lower.65ex\hbox{$\sim$}}\;}
\def\gsim{\mathrel{\gtap}}
\def\lsim{\mathrel{\ltap}}
\def\slashchar#1{\setbox0=\hbox{$#1$}           
   \dimen0=\wd0                                 
   \setbox1=\hbox{/} \dimen1=\wd1               
   \ifdim\dimen0>\dimen1                        
      \rlap{\hbox to \dimen0{\hfil/\hfil}}      
      #1                                        
   \else                                        
      \rlap{\hbox to \dimen1{\hfil$#1$\hfil}}   
      /                                         
   \fi}                                        %

\def\singleandabitspaced{\baselineskip=\normalbaselineskip\multiply
    \baselineskip by 120\divide\baselineskip by 100}
\def\singlespaced{\baselineskip=\normalbaselineskip}

\newcommand{\newc}{\newcommand}
\newc{ \nmess         }{n}
\newc{ \G         }{\tilde G}
\newc{ \Ni         }{ {\tilde N}_i }
\newc{ \Nj         }{ {\tilde N}_j }
\newc{ \NI         }{ {\tilde N}_1 }
\newc{ \NII        }{ {\tilde N}_2 }
\newc{ \NIII       }{ {\tilde N}_3 }
\newc{ \NIIII      }{ {\tilde N}_4 }
\newc{ \Ci         }{ {\tilde C}_i }
\newc{ \Cj         }{ {\tilde C}_j }
\newc{ \CI         }{ {\tilde C}_1 }
\newc{ \CII        }{ {\tilde C}_2 }
\newc{ \CIp        }{ {\tilde C}^{+}_1 }
\newc{ \CIpm       }{ {\tilde C}^{\pm}_1 }
\newc{ \CIm        }{ {\tilde C}^{-}_1 }
\newc{ \Cip        }{ {\tilde C}^{+}_i }
\newc{ \Cim        }{ {\tilde C}^{-}_i }
\newc{ \Cjp        }{ {\tilde C}^{+}_j }
\newc{ \Cjm        }{ {\tilde C}^{-}_j }
\newc{ \XI         }{ {\tilde X}_1 }
\newc{ \XII        }{ {\tilde X}_2 }
\newc{ \eL         }{ {\tilde e}_L }
\newc{ \eR         }{ {\tilde e}_R }
\newc{ \veL        }{ {\tilde \nu} }
\newc{ \SG         }{ {\tilde \gamma} }
\newc{ \SZ         }{ {\tilde Z} }
\newc{ \gmu        }{ \gamma^{\mu} }
\newc{ \gnu        }{ \gamma^{\nu} }
\newc{ \gamone      }{ \gamma_{1} }
\newc{ \gamtwo      }{ \gamma_{2} }
\newc{ \gamboth      }{ \gamma_{1,2} }
\newc{ \dL         }{ \tilde d_L }
\newc{ \dR         }{ \tilde d_R }
\newc{ \uL         }{ \tilde u_L }
\newc{ \uR         }{ \tilde u_R }
\newc{ \slepton    }{ \widetilde \ell }
\newc{ \M          }{ {\cal M} }
\newc{ \ra         }{ \rightarrow }
\newc{ \ltilde     }{ {\tilde \ell} }
\newc{ \nutilde    }{ {\tilde \nu} }
\newc{ \Qmess    }{ {Q_{\rm mess}} }
\newc{ \lLstar     }{ { \tilde \ell}_L^* }
\newc{ \lRstar     }{ { \tilde \ell}_R^* }
\newc{ \snu        }{ { \tilde \nu} }
\newc{ \stau        }{ { \tilde \tau} }
\newc{ \snustar    }{ { \tilde \nu}^* }
\newc{ \nubar      }{ \overline{ \nu } }
\newc{ \muL        }{ { \tilde \mu}_L }
\newc{ \muR        }{ { \tilde \mu}_R }
\newc{ \tauL       }{ { \tilde \tau}_L }
\newc{ \tauR       }{ { \tilde \tau}_R }
\newc{ \h          }{ { h^0 } }
\newc{ \Et         }{ { \slashchar{E}_T } }
\newc{ \Etot         }{ { \slashchar{E} } }
\newc{ \spt         }{ { \slashchar{p}_T } }
\newc{ \Etcut      }{ { \slashchar{E}_T^{\rm cut} } }
\newc{ \sigbreff   }{ \sigma \times {\rm BR} \times {\rm EFF} }
\newc{ \eegg       }{ {ee\gamma\gamma} }
\newc{ \GeV        }{ {\rm GeV} }
\newc{ \Dzero}{{D0}}

\newcommand{\sss}{\scriptscriptstyle}

\newcommand{\ser}{\tilde{e}_{\sss R}}

\newcommand{\smur}{\tilde{\mu}_{\sss R}}

\newcommand{\sneutrino}{\tilde{\nu}}

\newcommand{ \lRp }{ \tilde{\ell}_R^+ }
\newcommand{ \lRm }{ \tilde{\ell}_R^- }
\newcommand{ \ph  }{ \gamma }
\newcommand{ \E   }{ \slashchar{E} }

\begin{document}
\begin{titlepage}
\begin{flushright}
{\large
 hep-ph/9703211
 }
 \end{flushright}
\vskip 1.2cm

\begin{center}
{\LARGE\bf Signals for gauge-mediated supersymmetry breaking}

{\LARGE\bf models at the CERN LEP2  collider}

\vskip 2cm

{\large
 S.~Ambrosanio\footnote{{\tt ambros@umich.edu}},
 Graham D.~Kribs\footnote{{\tt kribs@umich.edu}}, and
 Stephen P.~Martin\footnote{{\tt spmartin@umich.edu}}
} \\
\vskip 4pt
{\it Randall Physics Laboratory, University of Michigan,\\
     Ann Arbor, MI 48109--1120 } \\

\vskip 1.5cm

\begin{abstract}

We consider a general class of models with gauge-mediated supersymmetry
breaking in which the gravitino is the lightest supersymmetric particle.
Several qualitatively different scenarios arise for the phenomenology of such
models, depending on which superpartner(s) decay dominantly to the gravitino.
At LEP2, neutralino pair production and slepton pair production can lead to 
a variety of promising discovery signals, which we systematically study.
We investigate the impact of backgrounds for these signals and show how they
can be reduced, and outline the effects of model parameter variations on the
discovery potential.

\end{abstract}

\end{center}

\vskip 3.0 cm

\end{titlepage}
\setcounter{footnote}{0}
\setcounter{page}{2}
\setcounter{section}{0}
\setcounter{subsection}{0}
\setcounter{subsubsection}{0}

\singleandabitspaced
\section*{I. Introduction}
\indent

Low-energy supersymmetry (SUSY) can provide a natural solution to the
hierarchy problem associated with the ratio $M_Z/M_{\rm Planck}$. If
nature is indeed supersymmetric, it is important to understand the mechanism
by which SUSY breaking occurs and is transmitted to the particles
of the standard model and their superpartners.
One possibility is that SUSY is broken at a scale $\sim10^{11}$ GeV
in a sector
which communicates with the particles of the Minimal Supersymmetric Standard
Model (MSSM) only through gravitational interactions. This has historically
been the most popular approach, and its phenomenological consequences have
been and continue to be well-studied. In this paper, we will be concerned
instead with
a different class of ``gauge-mediated SUSY breaking" (GMSB) models, in which
the messengers of supersymmetry breaking are the ordinary gauge interactions
\cite{oldmodels,GaugeMediated}.

Because gauge interactions are flavor-blind, GMSB models are highly
predictive with respect to the form of soft SUSY-breaking interactions.
In the minimal model of GMSB \cite{GaugeMediated},
the squark, slepton, neutralino,
and chargino masses are determined by only a handful of free parameters.
The MSSM gaugino mass parameters necessarily have a common complex phase, which
can be rotated away. Squarks and sleptons with the same electroweak quantum
numbers are automatically degenerate in mass, up to radiative corrections
involving Yukawa couplings which can be safely neglected for the sfermions
of the first two families. Thus GMSB models have the pleasant feature that
they are automatically free of excessive non-Standard Model
flavor-changing neutral currents; this also holds in a large class
of extensions and variations of the minimal model \cite{DGP1}-\cite{Borzumati}.
Furthermore, the sparticle mass pattern
is highly constrained even in extensions of the minimal model which
contain many more parameters. This means that sparticle spectroscopy may one
day provide for critical tests of GMSB.

However, the most distinctive phenomenological feature of GMSB models
may be that,
unlike in gravity-mediated SUSY breaking models, the
gravitino $(\G)$ is generally the lightest supersymmetric particle (LSP).
This is because the scale $\sqrt F$ associated with dynamical SUSY breakdown
can be as low as 10 TeV. The spin-3/2 gravitino obtains its
mass by the super-Higgs
mechanism, absorbing the spin-1/2 would-be Goldstino which couples to the
divergence of the supercurrent with strength $1/F$. The resulting gravitino
mass is
\beq
m_{\G} = {F\over {\sqrt{3}} M} = 2.37 \times 10^{-2}
\left ( {\sqrt{F}}\over 10\> {\rm TeV} \right )^2 \> {\rm eV}
\label{gravmass}
\eeq
where $M = (8 \pi G_{\rm Newton} )^{-1/2} = 2.4 \times 10^{18}$ GeV.
The next-to-lightest supersymmetric
particle (NLSP) can therefore decay into its standard model partner and
a gravitino \cite{Fayet,decay,DDRT}.
(In this paper we assume exact $R$-parity conservation, so that
there are no competing decays available for the NLSP.)
If 
the scale $\sqrt{F}$ does not exceed a few thousand TeV, the decays
can occur within a collider detector volume, possibly with a measurable
decay length. Furthermore, even supersymmetric particles which are
not the NLSP can decay into their standard model partners and a gravitino,
if no competing decays are kinematically allowed. As we will see below, this
may be
an important consideration. The perhaps surprising
relevance of a light gravitino for collider physics
\cite{Fayet}-\cite{DDN} can be traced to the fact that
the interactions of the longitudinal components of the gravitino
are the same as that of the Goldstino it has absorbed, and are proportional
to $1/m_{\tilde G}^2$ (or equivalently to $1/F^2$) in the light gravitino
(small $F$) limit \cite{Fayet}.

In the GMSB models to be considered in this study, the NLSP is always either a
neutralino or a charged slepton. In the former case, the lightest neutralino
($\NI$) decays into a photon and a gravitino with a width
\beq
& &\Gamma (\NI \rightarrow \gamma \G ) =
{\kappa_{1\gamma} \over 48 \pi}{m_{\NI}^5 \over M^2 m_{\G}^2} =
20\> \kappa_{1\gamma} \left ({m_{\NI}\over 100\>\rm{ GeV}}\right )^5
\left( {\sqrt{F}\over 10\>{\rm TeV}} \right )^{-4}
\>{\rm eV}
\label{neutralinodecaywidth}
\eeq
where
$\kappa_{1\gamma} = |N_{11}\cos\theta_W + N_{12}\sin \theta_W |^2$
is the photino component of $\NI$ (using the notation of \cite{HaberKane}
for the neutralino mixing matrices $N_{ij}$). The probability that an $\NI$
with
energy $E$ in the lab frame will decay before traveling a distance $x$ is
then
\beq
P(x) = 1 - e^{-x/L},
\label{probability}
\eeq
 where
\beq
L = 9.9\times 10^{-7} \> {1\over \kappa_{1\gamma}}
\left ({m_{\NI}\over 100\>\rm{ GeV}}\right )^{-5}
\left( {\sqrt{F}\over 10\>{\rm TeV}} \right )^4
\left ( E_{\NI}^2/m_{\NI}^2 -1 \right )^{1/2} \> {\rm cm}.
\label{neutralinodecaylength}
\eeq
In principle, one can also have $\NI \rightarrow Z\G$ or $h\G$,
but the corresponding decay widths \cite{AKKMM2}
suffer a strong kinematic suppression
and can easily be shown to be always negligible within the context of the
present paper.

In the rest frame of the decaying $\NI$, the photon is produced isotropically
(independent of the spin of $\NI$)
with energy equal to $m_{\NI}/2$.
The gravitino still escapes the detector, carrying away missing energy.
Therefore SUSY discovery signals at colliders
involve up to two energetic photons
and missing (transverse) energy in GMSB models
with a neutralino NLSP \cite{Fayet}-\cite{GKM}. At the Tevatron, the largest
production cross sections typically involve chargino ($\Ci$) and neutralino
($\Ni$) production, especially $p\overline p \rightarrow \CIp\CIm$ and
$\CIpm\NII$. One can therefore detect supersymmetry using an inclusive
$\gamma\gamma + \Et + X$ signal, in addition to channels with lepton(s) +
jet(s) + 0 or 1 photon. The discovery signatures for SUSY with a prompt
decay $\NI \rightarrow\gamma\G$ are so spectacular that it is possible
to set quite significant bounds even with existing Tevatron data.
For example, in \cite{AKKMM2}
it was argued that with the present $\sim 100$ pb$^{-1}$
of  Tevatron data, it should be possible to exclude chargino masses up to
about 125 GeV and neutralino masses up to about 70 GeV in a large class
of models with a light gravitino, including GMSB models, as long as the
decay $\NI\rightarrow\gamma\G$ occurs within the detector.\footnote{A quite
similar bound on $m_{\NI}$ in a large class of GMSB
models may be obtained using the recent
LEP runs at $\sqrt{s} = 161, 172$ GeV, as we will see in section III.}
In ref.~\cite{BBCT}, a significant reach was found
also in the 0 and 1 photon channels.

In discussing exclusion possibilities at the Tevatron, we must mention that
a single unusual event \cite{Event} of this general type has been observed
at CDF. This event has large ($>50$ GeV) $\Et$ and two central ($|\eta |< 1$),
energetic ($E_T>30$ GeV)
photons and two energetic leptons.
Events with these characteristics are reputed to have very small
Standard Model and detector backgrounds, and it was pointed out in
\cite{DDRT,AKKMM1} that this event might be explained by GMSB models (and
other models with a light gravitino) in terms of selectron pair production.
However, at least in the simplest types of GMSB models,
this interpretation is now perhaps somewhat disfavored,
since one might typically
expect many accompanying events in other channels
\cite{DTW,AKKMM2,BKW,BBCT},
which have not turned
up in recent searches by CDF \cite{CDFsearch} and
by \Dzero~\cite{Dzerosearch}.
Perhaps a more
plausible explanation of the event within the GMSB framework is that it
was due to chargino pair production $p\overline p \rightarrow \CIp\CIm$,
as proposed in \cite{AKKMM2,BKW}.
Each of the charginos can decay either
hadronically into $q\overline q^\prime\gamma\G$ or
leptonically into $\ell\nu\gamma\G$. The latter possibility can
be significantly enhanced if sneutrinos are not too heavy (although it
still seems somewhat problematic to explain the kinematics of the observed
event). GMSB models can be constructed with a neutralino NLSP
and a large leptonic branching fraction for $\CI$
(with $m_{\CI}>m_{\sneutrino}$),
but not with a minimal messenger sector.
The signal for $\CIp\CIm$ production in this case can then be
$\ell^+\ell^{\prime -}\gamma\gamma\Et$.
Note that in this chargino interpretation
of the event, the leptons need not have the same flavor.
In any case, more data at the Tevatron and at LEP2
will help to test these speculations.

In other GMSB models, one finds that the NLSP is a stau. Here, one should
distinguish between several qualitatively distinct situations. If $\tan\beta$
(the ratio of Higgs expectation values
$\langle H^0_u\rangle/\langle H^0_d\rangle$)
is not too large, the lightest stau eigenstate $\stau_1$
is predominantly right-handed
and is nearly degenerate in mass with the other right-handed sleptons.
In the case that $m_{\ser}\approx m_{\smur} < m_{\stau_1} + m_\tau$, one
finds that the $\ser$ and $\smur$ cannot have three-body decays
into $\stau_1$
without violating lepton flavor conservation.
Since
lepton flavor-changing interactions are automatically
very highly suppressed in GMSB models with $R$-parity conservation and
decays through an off-shell tau are insignificant,
each of the right-handed sleptons decays only
into the corresponding lepton + gravitino \cite{DDRT},
and $\stau_1$, $\smur$, $\ser$
act effectively as co-NLSPs. (An exception occurs if $m_{\stau_1} < m_{\NI} <
m_{\ser}$, as discussed below.) In this case, all supersymmetric decay
chains will terminate in $\stau_1 \rightarrow \tau\G$ or
$\ser \rightarrow e\G$ or $\smur \rightarrow \mu\G$.
The formulas for the
relevant decay widths of slepton into lepton + gravitino
are given by
simply replacing $m_{\NI}\rightarrow
m_{\slepton}$, $E_{\NI} \rightarrow E_{\slepton}$
and $\kappa_{1\gamma}\rightarrow 1$ in
(\ref{neutralinodecaywidth}) and (\ref{neutralinodecaylength}),
so that
the decay length in the lab frame for a slepton with energy $E_{\slepton}$ is
\beq
L = 9.9\times 10^{-7}\>
\left ({m_{\slepton}\over 100\>\rm{ GeV}}\right )^{-5}
\left( {\sqrt{F}\over 10\>{\rm TeV}} \right )^4
\left ( E_{\slepton}^2/m_{\slepton}^2 -1 \right )^{1/2} \> {\rm cm} .
\label{sleptondecaylength}
\eeq
At LEP2 energies, the primary discovery process often (but certainly
not always, as we shall see) involves  simple pair production
of the NLSP. In that case, $E_{\NI}$ in eq.~(\ref{neutralinodecaylength})
and $E_{\slepton}$ in eq.~(\ref{sleptondecaylength})
can simply be replaced by the electron beam energy at LEP2.

Conversely, if $\stau_1$ is the NLSP and $\tan\beta$ exceeds 4 to 8
(depending on the other model parameters), one finds in GMSB models that
$m_{\stau_1}$ is small enough 
that the decays $\slepton_R \rightarrow \stau_1
\tau\ell$ are kinematically allowed for $\ell = e,\mu$.
These three-body decays are mediated by virtual neutralinos
and are typically not dynamically
suppressed, because the bino content of $\NI$ is
significant. However, we have checked that
they can be quite strongly suppressed by phase space,
so that it is possible for $\slepton_R \rightarrow \ell\G$ to dominate
even
if $m_{\slepton_R} - m_{\stau_1} - m_{\tau} - m_{\ell}$ is a few GeV,
if $\sqrt{F}$ is not too large and $\slepton_R \rightarrow \ell\NI$ is
kinematically disallowed.
Barring these circumstances, $\stau_1$ acts as the sole NLSP, and
all supersymmetric
decay chains will terminate in $\stau_1\rightarrow\tau \G$ \cite{DTW2}.

An important exception to the preceding discussion occurs if
$|m_{\NI} - m_{\stau_1}| < m_{\tau}$ and $m_{\NI} < m_{\slepton_R} + m_\ell$
for $\ell = e,\mu$. Then each of the decays $\NI\rightarrow\gamma\G$ and
     $\stau_1 \rightarrow \tau\G$ have no significant
competition, and $\NI$ and $\stau_1$ act effectively as co-NLSPs.

To summarize, there are four qualitatively distinct scenarios for the
termination of supersymmetric decay chains
in GMSB models. By a slight abuse of language, we refer to these
as ``neutralino NLSP",
``stau NLSP", ``slepton co-NLSP", and ``neutralino-stau co-NLSP" scenarios,
according to whether A only, B only, B and C, or A and B of
the decays
\beq
A)& & \NI\rightarrow \gamma\G\\
B)& & \stau_1\rightarrow \tau\G\\
C)& & \slepton_R\rightarrow \ell\G\qquad(\ell=e,\mu)
\eeq
do not suffer competition.
The four possible scenarios correspond nominally to the mass orderings
(in addition to $m_{\stau_1} < m_{\slepton_R}$ for $\ell=e,\mu$, which turns
out to be always satisfied in the GMSB parameter space we consider):
\beq
\mbox{neutralino NLSP:}& &\>\>\> m_{\NI} < m_{\stau_1} - m_{\tau}
\label{codeone}\\
\mbox{stau NLSP:}& &\>\>\> m_{\stau_1} < {\rm Min}[m_{\NI},m_{\slepton_R}]
- m_{\tau}
\label{codetwo}\\
\mbox{slepton co-NLSP:}& &\>\>\>
m_{\slepton_R} < {\rm Min}[m_{\NI}, m_{\stau_1} + m_{\tau}]
\label{codethree}\\
\mbox{neutralino-stau co-NLSP:}& &\>\>\> |m_{\stau_1} - m_{\NI}|
< m_{\tau}
;\>\>
\>\>
m_{\NI} < m_{\slepton_R},
\label{codefour}
\eeq
where we have neglected the masses of the electron and the muon.
We should note that the condition eq.~(\ref{codetwo}) for the stau
NLSP scenario is necessary but not quite sufficient, since as we have already
noted, the decay $\slepton_R \rightarrow \ell\G$ can dominate over
$\slepton_R \rightarrow \stau_1\tau\ell$ when the latter is kinematically
open but suppressed. Thus some models which obey eq.~(\ref{codetwo}) 
may actually behave as slepton co-NLSP models, depending on $\sqrt{F}$.
We have checked that two-body decays $\slepton\rightarrow\ell\NI$ always
dominate over decays into the gravitino as long as $\sqrt{F} > 10$ TeV
and the mass difference
$m_{\slepton} - m_{\NI} - m_{\ell}$ is more than of order 10 MeV for
$\ell=e$ (and much less for $\ell=\mu,\tau$). A similar statement holds
for two-body decays $\NI \rightarrow \ell\slepton$. Hence we will consider
only the four main scenarios listed above. Other ``borderline" 
cases with small
mass differences will have similar phenomenology to the cases we do treat.

In section II of this paper we describe the framework for
an exploration of the parameter space of GMSB models, with some simplifying
but hopefully not overly restrictive assumptions.
In section III, we give some
conditions on the parameters for each of the
four different NLSP scenarios, and study the possible
signals and backgrounds
which arise at LEP2  in each case. We will mostly consider
an option with $\sqrt{s} = 190$ GeV and 300 pb$^{-1}$ per detector.
Section IV contains some concluding remarks.

\section*{II. Models of gauge-mediated SUSY breaking}\indent

In this paper we will consider the following class of GMSB models. The ultimate
source of SUSY breaking is parameterized by a gauge-singlet
chiral superfield $S$ whose
scalar and auxiliary components are both assumed to acquire
vacuum expectation values (VEVs), denoted
$S$ and $F_S$ respectively. The superfield $S$ couples to a ``messenger sector"
consisting of chiral superfields $\Phi_i,\overline\Phi_i$ which transform as
a vector-like representation of $SU(3)_C\times SU(2)_L \times U(1)_Y$.
The messenger sector couples to the SUSY-breaking sector through the
superpotential
\beq
W = \lambda_i S \Phi_i \overline\Phi_i.
\label{superpotential}
\eeq
This implies that the fermionic messengers acquire a Dirac mass $\lambda_i S$,
while their scalar partners obtain
(mass)$^2 = |\lambda_i S|^2 \pm |\lambda_i F_S|$. The ordinary gauge
interactions
then transmit this SUSY violation to the MSSM fields, with
computable superpartner masses \cite{oldmodels,GaugeMediated}.
Contributions to gaugino masses
due to each messenger pair $\Phi_i, \overline\Phi_i$
arise at one loop and are given at the scale(s)
$Q = \lambda_i S$ by
\beq
\Delta M_a = {\alpha_a\over 4\pi} \Lambda
n_a(i) g(x_i) \qquad\qquad (a=1,2,3)
\label{gauginomasses}
\eeq
where
\beq
&&\Lambda \equiv F_S/S,\\
&&g(x) =
{1\over x^2}[{(1+x)\ln (1+x) } +
{(1-x)\ln (1-x) }]\> ,
\eeq
and each
\beq
x_i \equiv |F_S/\lambda_i S^2|.
\eeq
The latter quantities must satisfy
$0<x_i<1$ (so that the lightest messenger scalar
does not acquire a VEV). Here $n_a(i) $ is the Dynkin index for the
messenger pair $\Phi_i,\overline\Phi_i$ in a normalization where $n_a=1$ for
${\bf N} + {\bf \overline N}$ of $SU(N)$. We always use a
Grand Unified Theory (GUT)
normalization for $\alpha_1$ so that $n_1 = {6\over 5} Y^2$ for
each messenger pair with weak hypercharge $Y=Q_{\rm EM} - T_3$.
In the limit of small $x_i$ and when $\Phi_i,\overline\Phi_i$ consist of a
${\bf 5} + \overline{\bf 5}$ of the global $SU(5)$ which contains
$SU(3)_C \times SU(2)_L \times U(1)_Y$, one recovers the results for
the original minimal GMSB models with $\sum n_a = 1$ and small $x_i$
\cite{GaugeMediated}, since $g(0) = 1$. The function $g(x)$ is always
slightly greater than 1, but never exceeds 1.044 when $x<0.5$,
and reaches a maximum of
1.386 at $x=1$ \cite{GernBlanston}.

Contributions from each messenger pair $\Phi_i,\overline\Phi_i$
to the (mass)$^2$ terms of the MSSM scalars arise
at two-loop order and are given at the scale(s) $Q=\lambda_i S$ by
\beq
\Delta {\tilde m}^2=
2 \Lambda^2 \sum_a \left ({\alpha_a\over 4\pi}
\right )^2 C_a n_a(i) f(x_i)
\label{scalarmasses}
\eeq
where \cite{DGP2}
\beq
f(x) =
{1+x\over x^2}\biggl [ \ln (1+x) - 2 {\rm Li}_2(x/[1+x])  +
{1\over 2} {\rm Li}_2(2x/[1+x])\biggr ]
+ (x \rightarrow -x)
\label{definef}
\eeq
and  $C_a$ is the quadratic Casimir invariant
of the MSSM scalar field in question, in a normalization where
$C_3=4/3$ for color triplets, $C_2 = 3/4$ for $SU(2)_L$ doublets,
and $C_1 = {3\over 5} Y^2$.
For small $x_i$, one has $f(x_i) \approx 1$, so that again
the results of the original minimal model
with small $x_i$ \cite{GaugeMediated} are recovered.

In order to have a manageable parameter space for our study, we
now make some simplifying assumptions. First,  we consider (except when
explicitly noted) only models
for which the total Dynkin indices of the messenger sector for each
gauge group are equal and
do not exceed 4:
\beq
n = \sum_i n_1(i) = \sum_i n_2(i)
= \sum_i n_3(i) = 1,2,3,\>{\rm or}\> 4.
\label{defnmess}
\eeq
This assumption ensures that the apparent near-unification of
perturbative gauge couplings near $M_U \approx 2\times 10^{16}$ GeV
is maintained. (Possibilities which do not embrace this assumption are
discussed in \cite{GernBlanston}.)
We will also take all of the couplings $\lambda_i$
to be equal to a common value $\lambda$, even though no symmetry can enforce
this; variations in the individual $\lambda_i$ only affect the
MSSM sparticle spectrum logarithmically.
This in turn forces all of the $x_i$ to be equal to a single
parameter $x$, which as a practical matter we require to satisfy
\beq
0.01 < x < 0.9\>.
\label{xconstraint}
\eeq
With these assumptions, MSSM phenomenology is determined by just 6 parameters
\beq
\Lambda,\>\> n,\>\> x, \>\>\lambda,\>\> \tan\beta,\>\>{\rm and}
\>\>{\rm sign}(\mu)\>.
\label{parameters}
\eeq
The expressions for the sum of contributions to
gaugino and sfermion masses now simplify to
\beq
M_a = n \Lambda g(x) {\alpha_a \over 4 \pi}
\label{simplegauginomasses}
\eeq
\beq
\tilde m^2 = 2 n \Lambda^2 
f(x) \sum_a \left ({\alpha_a \over 4 \pi}\right )^2 C_a
\label{simplesfermionmasses}
\eeq
at the single messenger scale $\Qmess = \Lambda/x$.
Equations (\ref{simplegauginomasses}) and
(\ref{simplesfermionmasses}) are taken as boundary conditions
for renormalization group (RG)
evolution of the MSSM parameters. At the same scale $\Qmess$,
the running trilinear scalar couplings of the MSSM are taken to vanish
(they actually receive contributions at two-loop order which are negligible
in the first approximation). However, we do not assume that $B\mu$ is close to
zero at the messenger scale, since it seems
likely that a mechanism for generating
$\mu$ can also generate
$B\mu$ \cite{GaugeMediated,DGP1,DNS}. We then evolve
all of the couplings of the MSSM
from $\Qmess$ down to the electroweak scale, where the parameters $B\mu$
and
$|\mu |$ are determined by requiring correct electroweak symmetry breaking.
Note that in this parameterization, the sparticle spectrum (with the
exception of the gravitino mass!) does not depend on $\lambda$, and depends
on $x$ only logarithmically through $\Qmess$ and $g(x)$.
Of course, allowing different $x_i,\lambda_i$ would be more realistic
and would enlarge the parameter space. However, the features of the enlarged
parameter space obtained in this way do not differ dramatically from
the one we consider. The effect of finite $x_i$ in
(\ref{gauginomasses}) and (\ref{scalarmasses})
is simply to raise the gaugino mass parameters by up to about 25\% 
with respect to the
sfermion masses, since $g(x)/\sqrt{f(x)}$ varies between 1 (for $x\ll 1$)
and 1.25 (for $x=0.9$). Choosing a specific value of $x$ can be
thought of as simply parameterizing our ignorance of these effects
within a simplified framework. In the present paper, the chargino,
squark, and
gluino masses have no direct relevance, so that the practical effect
of increasing $x$ is essentially just to lower the slepton masses compared to
the mass of the lightest neutralino.

The Goldstino decay constant in this parameterization is given by
\beq
F_S = {\Lambda^2 \over x_i \lambda_i} = {\Lambda^2 \over x \lambda}
\label{Fexp}
\eeq
This way of expressing $F_S$ is useful because $\Lambda$ is relatively
well-known from
(\ref{simplegauginomasses}) and (\ref{simplesfermionmasses}), since it is
correlated strongly with the mass of the NLSP which we must presume
lies in the 50 to 100 GeV range in order for SUSY to be relevant at LEP2.
An easy estimate then shows that the relevant range of $\Lambda$ for
this paper is from about 10 TeV to about 100 TeV.
While the dimensionless couplings $x$ and $\lambda$ can be arbitrarily small,
they can be bounded from above.
Furthermore, 
$F_S$ can be smaller than the full SUSY-breaking order parameter
$F$ of the complete theory. (In models where $F_S$ arises directly from
an O'Raifeartaigh mechanism one expects $F_S=F$, but in models where
$F_S$ itself arises radiatively as in \cite{GaugeMediated}, one may
find $F_S\ll F$.)
Therefore eq.~(\ref{Fexp}) puts a lower limit on $F$, or equivalently a
lower limit on $m_{\tilde G}$, corresponding to a lower limit on the
decay length of the NLSP according to eqs.~(\ref{neutralinodecaylength})
and (\ref{sleptondecaylength}).
In particular, we note that in viable GMSB models one must have
\beq
m_{{\G}} \gsim 2\times 10^{-2}\>{\rm eV},
\eeq
based only on $\lambda_i \lsim 1$ and $m_{\rm NLSP} > 50$ GeV; this is
consistent with cosmological bounds \cite{cosmoconstraints} on $m_{{\G}}$.
For any given sparticle spectrum specified
by the parameters $\Lambda,\nmess,x,\tan\beta,{\rm sign}(\mu)$, one obtains
a lower limit on the NLSP decay length by assuming $\lambda\lsim 1$.
By
taking smaller $\lambda$ with $\Lambda$ and $x$ held constant
one can essentially arbitrarily increase the NLSP decay length
while holding all other features of the MSSM sparticle spectrum fixed.
The NLSP decay length will also be increased if $F_S<F$, as long as
a pseudo-Goldstino field which is not absorbed by the gravitino (and which is
predominantly the fermionic component of $S$) acquires a large mass.
In these cases the NLSP decay length $L$ can be made so large that NLSP
decays always occur outside the detector.

The statement that $\lambda$, or more generally the distinct
couplings $\lambda_i$, should be bounded from above can be
motivated as follows.
An estimate of the maximum value of the couplings $\lambda_i$
should roughly correspond to an infrared quasi-fixed point of the RG equations.
Let us consider, for example, a ``minimal" messenger sector in
which $\Phi_i,\overline\Phi_i$ consist of a ${\bf 5 } + {\bf \overline 5}$
of the global $SU(5)$ group which contains the MSSM gauge group. In
that case one has couplings $\lambda_2$ and $\lambda_3$ of $S$ to the
$SU(2)_L$-doublet and $SU(3)_C$-triplet messenger fields, respectively.
These couplings satisfy the one-loop RG equations
\beq
& &16 \pi^2 {d\lambda_2\over dt} = \lambda_2\left[ 4 \lambda_2^2 +
3 \lambda_3^2 + \ldots - 3 g_2^2 - {3\over 5} g_1^2 \right ] \\
& &16 \pi^2 {d\lambda_3\over dt} = \lambda_3\left[ 2 \lambda_2^2 +
5 \lambda_3^2 + \ldots - {16\over 3} g_3^2 - {4\over 15} g_1^2  \right ] \> .
\eeq
Here the ellipses represent the effects of other dimensionless
couplings
involving $S$, $\Phi_i$, or $\overline\Phi_i$.
These are of course highly model-dependent
but will contribute positively to the one-loop $\beta$ functions, thus only
reducing the quasi-fixed point values of the couplings.
One can then estimate the maximum
quasi-fixed point values for
$(\lambda_2,\lambda_3)$ by taking
$\lambda_3 = \lambda_2 = \lambda_U$ to be large
at the putative unification scale
and evolving down to the messenger scale. For
example if the messenger scale is $Q_{\rm mess} = 100$ TeV,
then taking $\lambda_U \gsim 2$
yields $(\lambda_2,\lambda_3) \approx (0.7,\>1.1)$ at $\Qmess$.
Even if we abandon GUT boundary conditions in this example and choose
$\lambda_3\ll\lambda_2$ or $\lambda_2\ll\lambda_3$ at the
``unification" scale, the maximum values at the messenger scale
of the dominant coupling are found to be
$\lambda_2 \approx 1.0$ or
$\lambda_3\approx 1.2$, respectively.
Of course, $S$ need not be a fundamental chiral superfield, but the
fixed point values for $\lambda_i$ should roughly correspond to the maximum
values, at least in a perturbative effective field theory description.
Note that models with more messenger fields should respect the rough result
$\lambda_i\lsim 1$. Adding
more couplings to the mix will effectively only give positive
contributions
to the corresponding $\beta$ functions when compared with
the example given above, leading to smaller values for the $\lambda_i$
at the messenger scale. A larger messenger sector will cause the gauge
couplings to be larger above the messenger scale, indirectly resulting in
a decrease in the $\beta$ functions for the $\lambda_i$. However, this can
only slightly increase the quasi-fixed point behavior of the
largest coupling(s), which is determined predominantly by
what happens near the messenger scale. Therefore we expect that the
lower bound on the NLSP decay length at LEP to be inferred from
$\lambda \lsim 1$ (for a given $x$) should be robust. In particular,
we see from (\ref{Fexp}) that for models with a given $\Lambda$, the minimum
possible decay length $L$ for the NLSP is given by replacing
$\sqrt{F} \rightarrow \Lambda$ in (\ref{neutralinodecaylength}) or
(\ref{sleptondecaylength}).

In order to understand the parameter space of the GMSB models we have
chosen to study,
we have used a computer program to generate several tens of thousands of models
for each of $\nmess=1,2,3,4$ and random values for the other free parameters in
eq.~(\ref{parameters}).
The program proceeds by an iterative method that sets the weak-scale
gauge couplings and masses, evolves the RG equations to the messenger scale,
sets the messenger scale boundary conditions, evolves the
RG equations with associated decouplings at each sparticle
threshold back to
the weak scale, then iterates to
convergence (about 4 iterations are typically necessary). Two-loop RG
equations are used for the gauge couplings, third generation Yukawa couplings,
and gaugino soft masses, while one-loop RG equations are used for the other
soft masses and scalar trilinear couplings. Electroweak symmetry breaking
is enforced at the scale $Q=\sqrt{m_{\tilde t_1} m_{\tilde t_2}}$, allowing
the evaluation of $|\mu|$ and $B\mu$ from the Higgs soft masses, $\tan\beta$,
$M_Z$, and one-loop corrections. This evaluation utilizes the one-loop
effective potential, which includes the corrections from stops, sbottoms,
and staus consistently with the one-loop evaluation of the Higgs masses.

Because this paper is devoted to possible discovery
signals at LEP2, we consider only models with NLSP mass less than 100 GeV.
The lightest supersymmetric particle is always a neutralino
or a stau throughout this parameter
space.\footnote{We find one exception:
it is possible to construct models with $n=3$ or 4 which have a tau sneutrino
NLSP, but {\it only} if
$m_{\tilde\nu_\tau} < 54$ GeV and $m_{\slepton_R} < 57$ GeV
and $m_{\slepton_L} < 95$ GeV.
We neglect this possibility in the following, although
a complete analysis which might exclude these models has not yet been performed
to our knowledge.}
However, as explained in the Introduction, more than one superpartner can
act effectively as the NLSP.
We find significant regions of parameter space in which each of
the four scenarios is indeed realized.
In the next section we will describe in turn some relevant
features of the parameter space, including conditions on
the parameters $\nmess$, $\Lambda$, and $\tan\beta$, for the
four NLSP scenarios.
In each case we study how the SUSY discovery signals may manifest themselves
at LEP2.

\section*{III. Signals at the CERN LEP2  collider}\indent
\subsection*{A. The neutralino NLSP scenario}
\indent

If equation (\ref{codeone}) is satisfied,
then the lightest neutralino is the NLSP, and essentially all
supersymmetric decay chains will terminate in $\NI\rightarrow\gamma\G$.
If this decay always occurs outside of the detector, then collider signatures
and search strategies are the same as in the well-studied neutralino LSP
scenario (see for example \cite{CDMT}, \cite{BBMT}).
In that case, the only impact of the gauge-mediation mechanism
for collider phenomenology is that the pattern of supersymmetric masses
and other soft supersymmetry breaking parameters is restricted in significant
ways.

If the decay $\NI\rightarrow\gamma\G$
occurs within the detector an appreciable fraction of the time, then
$\NI\NI$ production can lead to a discovery signal at LEP2.
A crucial quantity is then
the decay length $L$ in equation (\ref{neutralinodecaylength}) with
$E=\sqrt{s}/2$. For
$L$ greater than a few centimeters,
the LEP detectors may be able to resolve the distance
from the interaction point to the $\NI$ decay vertex where the photon
originates. If this can be done reliably, there should be
essentially no backgrounds to the signal. However, as $L$ increases, a larger
fraction of events will occur outside the detector, decreasing the
efficiency accordingly. Of course, any analysis of this situation would
be highly detector-dependent, as the different LEP detectors have varying
geometries and photon direction resolution capabilities. For a rough study
we suppose that photons resulting from decays $\NI \rightarrow \gamma\G$
can be detected if  they occur within 1 meter of the interaction point.
Using eqs.~(\ref{probability}),(\ref{neutralinodecaylength}),
we can then estimate, as a function of
the parameter $L$, the probabilities
that both, or only one, of the photons can be detected in each event.
\begin{figure}[tb]
\centering
\epsfxsize=3.8in
\hspace*{0in}
\epsffile{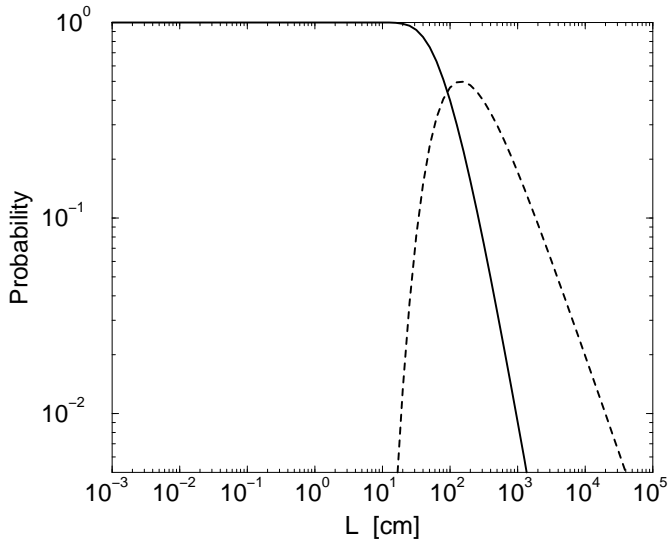}
\caption{The probability that exactly one (two) of the photons
from the decay $\NI\rightarrow\gamma\G$ originates within
1 meter of the $\NI\NI$ production vertex is shown as the dashed (solid)
curve, as a function of the decay length $L$.}
\label{figonephoton}
\end{figure}
These probabilities are shown in Fig.~\ref{figonephoton}
for $10^{-3}$ cm $< L < 10^5$ cm.
(This range corresponds roughly to $1 \gsim x \lambda \gsim 10^{-4}$
for the models described in section II with
$ m_{\NI} <$ 90 GeV and $\sqrt{s} = 190$ GeV.)
Note that the probability to observe one of the two photons in each event may
exceed 0.1 even for $L$ greater than 10 meters. Remarkably,
 an observable signal
with displaced (not originating from the interaction point)
single photons might occur for $L$ up to several tens of
meters, depending on the $\NI\NI$ production cross section (typically in
the tens or hundreds of fb),
the integrated luminosity achieved, and the specific detector
being used.
For $L$ in the several centimeter
to several meter range, one can hope to observe both one photon and two
photon events with displaced vertices.

For the remainder of this subsection, we assume that $L$ is less than
a few tens of centimeters, so
that essentially all decays $\NI\rightarrow\gamma\G$
occur within the detector.
We consider only models for which
$m_{\NI} < 100$ GeV (so that $\NI\NI$ production can be possible
at LEP2) and $m_{\NI} > 70$ GeV (motivated \cite{AKKMM2} by
the non-observation of excessive $\gamma\gamma+\Et+X$ events at
the Tevatron \cite{CDFsearch,Dzerosearch}).
Then we find that within the
framework of GMSB models assumed here,
neutralino NLSP models can only be constructed if
\beq
& &n=1; \qquad 40 \>{\rm TeV}\> < \> \Lambda < \> 80\> {\rm TeV};\qquad
\tan\beta < 35, \>\>{\rm or} \label{NNLSP1} \\
& &n=2; \qquad 24 \>{\rm TeV}\> < \> \Lambda < \> 40\> {\rm TeV};\qquad
\tan\beta < 18, \>\>{\rm or} \label{NNLSP2} \\
& &n=3; \qquad 20 \>{\rm TeV}\> < \> \Lambda < \> 24\> {\rm TeV};\qquad
\tan\beta < 10\>\>{\rm and}\>\>\mu > 0. \label{NNLSP3}
\eeq
These conditions are necessary, but not sufficient.
The upper and lower bounds on $\Lambda$
correspond to those on $m_{\NI}$ in the
obvious way. The upper limits on $\tan\beta$ follow from the
requirement that $m_{{\tilde\tau}_1} > m_{\NI} + m_\tau$;
for larger values of $\tan\beta$
the mixing in the stau (mass)$^2$ matrix becomes too large to allow this.
In all cases, one finds that $|\mu| \gsim M_2$, so that the NLSP
has a significant photino component, with
$0.4 < \kappa_{1\gamma} < 0.85$ in all models of this type, and
$\kappa_{1\gamma} > 0.55$ in the minimal model (meaning $\nmess=1$).
We have verified that
for all neutralino NLSP models accessible at LEP2, the BR$(\NI \rightarrow
\gamma\G)$ is in practice indistinguishable from 100\%.
In the models described above, the minimum possible decay length $L$
[estimated by substituting 
$E_{\NI}=\sqrt{s}/2$ and $\sqrt{F} \approx \Lambda$
in eq.~(\ref{neutralinodecaylength})] is typically of
order 10 to 100 microns, far smaller than the resolution of the detectors.

Neutralinos can be pair-produced in $e^+e^-$ collisions by $Z$ exchange
in the $s$-channel or by selectron exchange in the $t$-channel. In GMSB models,
$\NI$ is always predominantly a bino and the ratio $m_{\ser}/m_{\NI}$
cannot be larger than (1.6, 1.25, 1.1) for $\nmess=(1,2,3)$.
Therefore the dominant contribution
to $\NI\NI$ production always comes from $\ser$ exchange because of
the relatively large $e$-$\ser$-bino coupling. Indeed,
the non-minimal GMSB models with $\nmess=2,3$ tend to have a larger
cross section for $e^+e^- \rightarrow \NI\NI$ because of relatively lighter
$m_{\ser}$ for fixed $m_{\NI}$.
In addition, in some models $\slepton_R\slepton_R$ production (or just
$\stau_1\stau_1$) production is allowed at LEP2.
However, we find that in the neutralino NLSP scenario, the individual
slepton
production cross sections are always smaller by at least a factor of
2 (and often much more), so we will concentrate first
on $\NI\NI$ production as the discovery process.
We also find that chargino pair production
is never allowed at LEP2,
and $\NI\NII$ is sometimes allowed but is always highly
kinematically suppressed
and therefore insignificant. This is easily understood due to the
assumed bound $\NI \gsim 70$ GeV
and the rough relations $m_{\NII} \sim m_{\CI} \sim 2 m_{\NI}$, which hold
since $\mu$ is relatively large in our models.

The discovery process $e^+e^- \rightarrow
\NI\NI$ with each $\NI \rightarrow \gamma\G$, leads to events with two
energetic photons and large missing energy.\footnote{We prefer not
to use the words ``acoplanar photons" to refer to this signal, since
the acoplanarity seems not to be a particularly useful discriminant
against background.}
This signal has already been studied for LEP2  in an earlier paper
\cite{AKKMM2}, but we will be able to extend these results.
Also, ref.~\cite{AKKMM2} made no assumptions about model parameters;
in taking into account the constraints inherent in
the GMSB models we will be able to make some more
concrete (and optimistic!) statements.
The most important point to be made in this regard is that
in the GMSB neutralino NLSP models of the class described in section II,
the $\NI\NI$ production cross section can be bounded from below
for a given $m_{\NI}$. This result is due to the facts that $\NI$
is always gaugino-like and $m_{\ser}$ is bounded from
above. The dominant $t$-channel exchange of $\ser$ therefore always
ensures a substantial cross section. 
\begin{figure}[tb]
\centering
\epsfxsize=3.8in
\hspace*{0in}
\epsffile{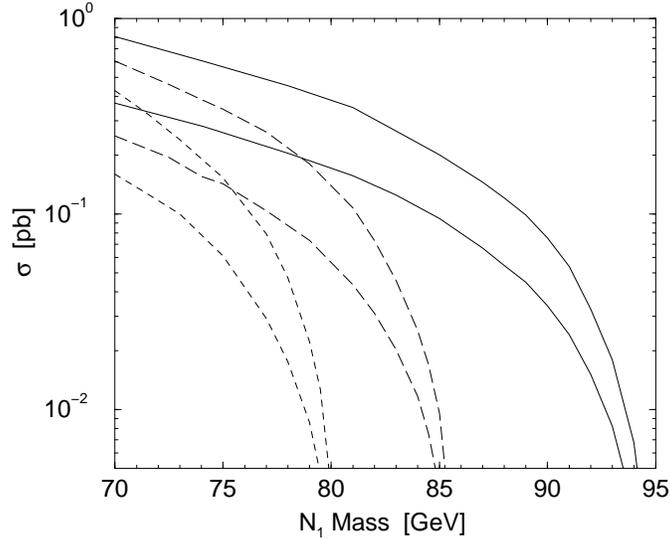}
\caption{The maximum and minimum cross sections for
$e^+e^-\protect\rightarrow\NI\NI$
at $\protect\sqrt{s}=161$ GeV (short-dashed), 172 GeV (long-dashed),
and 190 GeV (solid),
as a function of the mass of the lightest
neutralino. These bounds hold in the neutralino NLSP models
within the GMSB framework described in section II.}
\label{figminmax}
\end{figure}
To illustrate this, we show in
Figure \ref{figminmax} the minimum and maximum cross sections for $\NI$ pair
production obtained in these models, for $\sqrt{s} = 161$, 172, and 190 GeV.
This graph was prepared by an
exhaustive scan of the model parameter space, varying
$\Lambda$, $x$, $\tan\beta$, and sign($\mu$).
The minimum cross sections
are obtained for models with $\nmess=1$ and $x$ not too large
in which $m_{\ser}/m_{\NI}$ saturates
its upper bound of about 1.6.

The energy of each $\NI$ is equal to the beam energy ${\sqrt s}/2$, so
that the photon energies in each event have a flat distribution, with
\beq
E_{\rm min} < E_{\gamma_1}, E_{\gamma_2} < E_{\rm max}
\label{photonenergyrange}
\eeq
where
\beq
E_{\rm max, min} = {1\over 4}({\sqrt{s}} \pm \sqrt{s - 4 m_{\NI}^2})\> .
\label{photonenergybounds}
\eeq
Therefore one can always impose a cut on soft photons,
depending on an assumed lower bound on the mass of the $\NI$ being searched
for.
In this paper, we will (motivated by \cite{AKKMM2})
take $m_{\NI}> 70$ GeV as a given, so that a cut
\beq
E_\gamma > {1\over 4}({\sqrt{s}} - \sqrt{s -(140\>{\rm GeV})^2 })
\label{softphotoncut}
\eeq
on soft photons can be applied without affecting the signal at all.
The missing energy in each event is also bounded according to
$2 E_{\rm min} < \Etot < 2 E_{\rm max}$.

\begin{figure}[tb]
\centering
\epsfxsize=3.7in
\hspace*{0in}
\epsffile{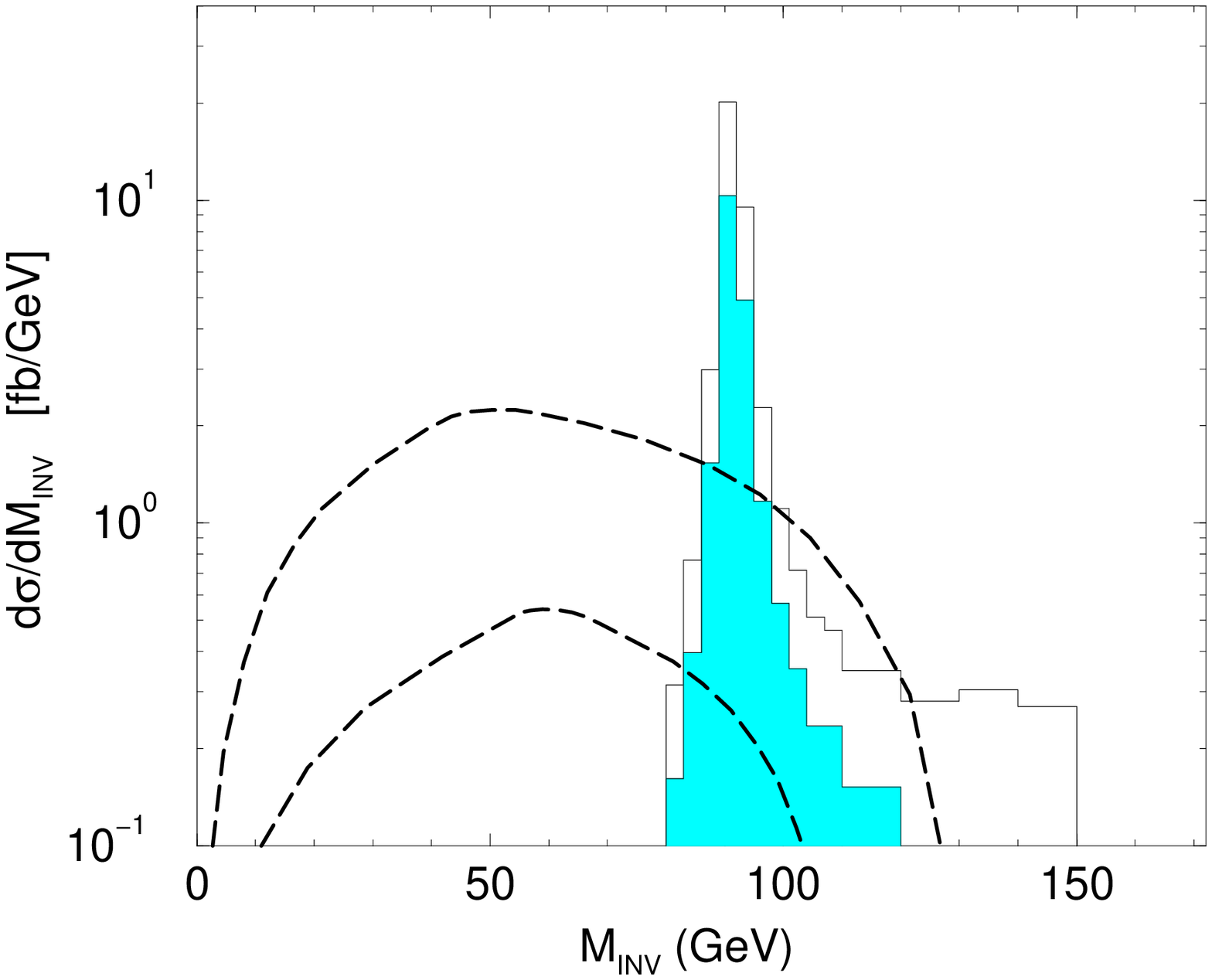}
\caption{Missing invariant mass distributions at $\protect\sqrt{s} = 172$ GeV
for $\gamma\gamma\E$ events which pass the cuts (\ref{photonanglecut}) and
(\ref{photonptcut}).
The background from $e^+e^-\rightarrow \gamma\gamma\nu_i\overline\nu_i$
is shown as the open (shaded) histogram before (after) the photon
energy cut $E_{\gamma} > $ 18 GeV from (\ref{softphotoncut}).
The upper (lower) dashed line
is the signal distribution from $\NI\NI$ production for representative
models with $m_{\NI} =$
75 (82) GeV as described in the text.}
\label{minv172fig}
\end{figure}
\begin{figure}[tb]
\centering
\epsfxsize=3.7in
\hspace*{0in}
\epsffile{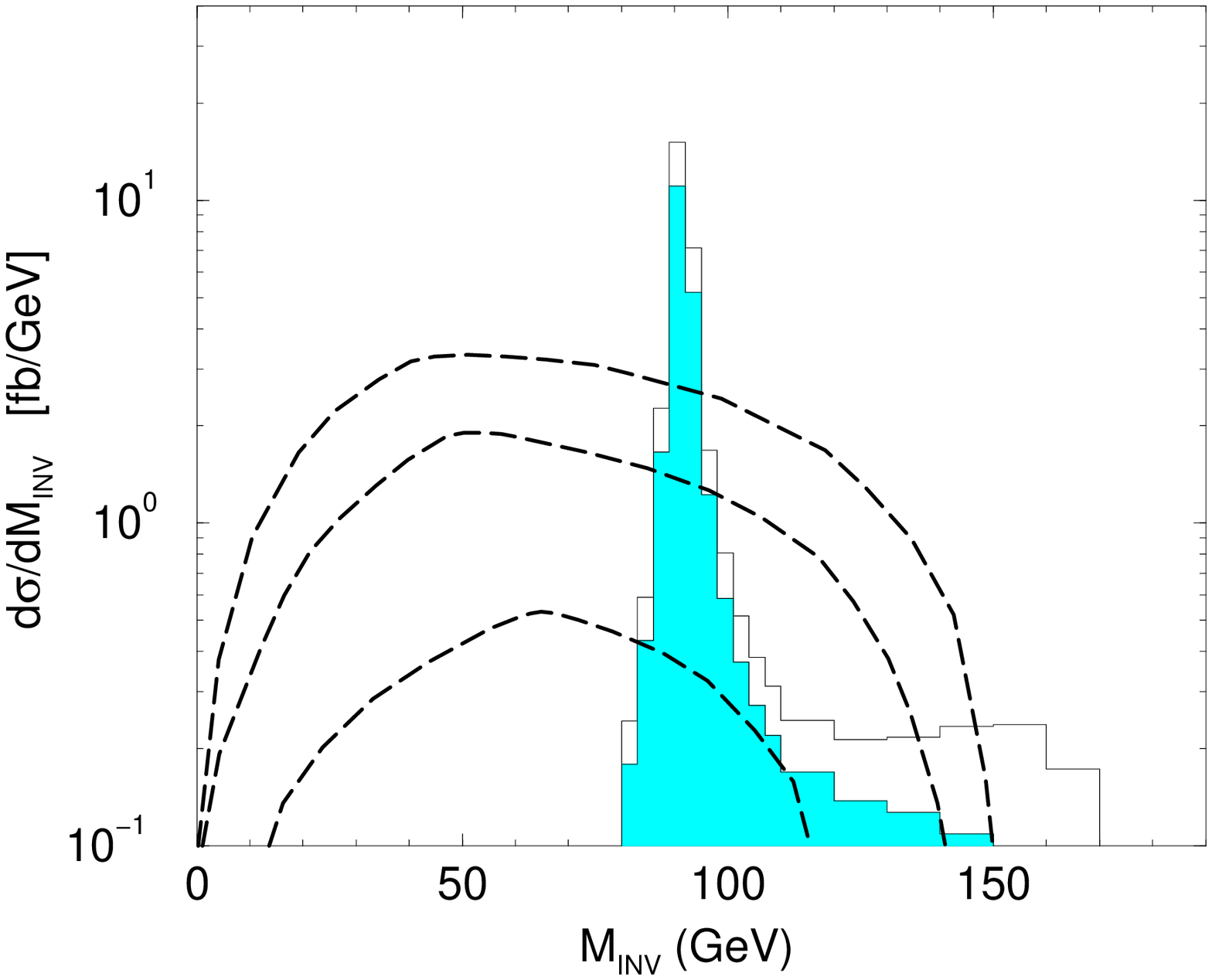}
\caption{Missing invariant mass distributions at $\protect\sqrt{s} = 190$ GeV
for $\gamma\gamma\E$ events which pass the cuts (\ref{photonanglecut}) and
(\ref{photonptcut}).
The background from $e^+e^-\rightarrow \gamma\gamma\nu_i\overline\nu_i$
is shown as the open (shaded) histogram before (after) the photon
energy cut $ E_{\gamma} > $ 16 GeV from (\ref{softphotoncut}).
The dashed lines
are the signal distributions from $\NI\NI$ production
for representative models with $m_{\NI} =$
75, 82, and 90 GeV (from top to bottom) as described in the text.}
\label{minv190fig}
\end{figure}

The most important physics backgrounds for the $\gamma\gamma\Etot$ signal
are due to $e^+e^- \rightarrow \gamma\gamma\nu_i{\overline\nu}_i$
from diagrams with
$s$-channel $Z$ exchange ($i=e,\mu,\tau$) and $t$-channel $W$ exchange
($i=e$ only). These backgrounds were discussed in some detail
in \cite{AKKMM2}, where it was shown that they can be
efficiently eliminated using a cut on the invariant missing mass
$M_{\rm INV}^2 \equiv (p_{e^+} + p_{e^-} - p_{\gamma_1} - p_{\gamma_2})^2$.
The $M_{\rm INV}$ distribution of the signal tends to be broadly distributed
and has most of its support
for lower values than the background, which is strongly peaked at $M_Z$
but with a significant tail due to the $t$-channel contributions.
In Figs.~\ref{minv172fig} and \ref{minv190fig} we show a comparison
of the signal and background distributions for $\sqrt{s} = 172$ and 190 GeV,
respectively. In preparing these figures, we have applied
detectability cuts \cite{detectorcuts}
\beq
& & |\cos \theta_\gamma
 | < 0.95\> ,
\label{photonanglecut}\\
& & (p_T)_\gamma
> 0.0325 \> \sqrt{s} 
{},
\label{photonptcut}
\eeq
for each photon. The background distributions before and after the
cut on soft photons (\ref{softphotoncut}) were computed at tree level using
CompHEP \cite{CompHEP}, and are shown as the
open and shaded histograms. 
Also shown are the distributions for the signals
derived at tree-level from some sample
models with $m_{\NI} =$ 75, 82, and 90 GeV. These models were chosen
with $\nmess=1$, $\tan\beta = 3$, $x=0.1$, and $\mu<0$. The shape (but
not the magnitude) of the
signal distributions is largely independent of these choices, for a fixed
$\sqrt{s}$ and $m_{{\NI}}$.
This can be easily understood since the dominant kinematic features of these
events are due to the isotropic decays $\NI\rightarrow\gamma\G$.
Besides the cuts eqs.~(\ref{softphotoncut})-(\ref{photonptcut})
we therefore propose to implement a
cut on the missing invariant mass of
\beq
5 \>\GeV \> < \> M_{\rm INV} \> < \>80\>\GeV \>.
\label{minvcut}
\eeq
The lower limit in (\ref{minvcut}) is designed to remove detector
backgrounds such as $e^+e^-\rightarrow\gamma\gamma(\gamma)$
with the third photon unobserved. We expect that the
     effect of initial state radiation will be to slightly reduce the
     total magnitude of both the signal 
     and the background,
     while not qualitatively affecting the
     shapes of $M_{\rm INV}$ distributions for our purposes.
While significant quantitative changes are possible
for $M_{\rm INV} \gsim m_Z$, the cut eq.~(\ref{minvcut}) makes
these effects irrelevant here.

At $\sqrt{s}=172$ GeV the total $\NI\NI$ production cross sections for
the examples in Fig.~\ref{minv172fig} with
$m_{\NI}=75$ and $82$ GeV are 195 fb and 41 fb respectively, before any
cuts.
After the cuts eqs.~(\ref{softphotoncut})-(\ref{minvcut}),
the remaining
background is less than 1 fb, while the
signal for these models is 128 fb and 30 fb.
With 10 pb$^{-1}$ per detector, this amounts to an expectation of
roughly 5 events and 1.2 events (summed over all four detectors).

At $\sqrt{s}=190$ GeV the total  cross sections in the models of
Fig.~\ref{minv190fig} are (350, 180, 41) fb for
$m_{\NI}=$ (75, 82, 90) GeV, respectively.
After applying all of the cuts eqs.~(\ref{softphotoncut})-(\ref{minvcut}),
we find
remaining signals of (193, 106, 25) fb. With 300 pb$^{-1}$ per detector,
this corresponds to about (58, 32, 7.5) events for each of the four detectors.
The remaining background after these cuts is less than 1 fb.

\begin{figure}[tb]
\centering
\epsfxsize=4in
\hspace*{0in}
\epsffile{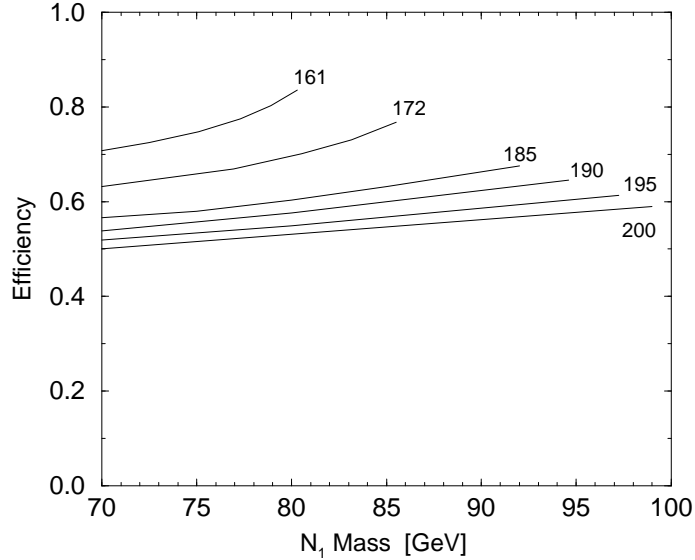}
\caption{The fraction of
$\protect{\gamma\gamma\Etot}$ events from neutralino pair
production which pass the
detector cuts (\ref{photonanglecut}), (\ref{photonptcut}),
and missing invariant mass cut (\ref{minvcut}). The lines
from top to bottom are for $\protect\sqrt{s} =$
161, 172, 185, 190, 195 and 200 GeV.
Here we have chosen a representative class of models as described in the
text, but for fixed
$m_{\protect \NI}$ and
$\protect\sqrt{s}$ the efficiency is
quite insensitive to variations in model parameters.}
\label{efficiencyfigure}
\end{figure}

As shown by these examples, the efficiency for detecting
signal events after cuts is quite high, while the physics background
is essentially completely eliminated.
The effects of the cuts on the backgrounds are shown in Table 1.
In Fig.~\ref{efficiencyfigure}, we show the efficiency
(defined as the fraction of signal events which pass all of our cuts
divided by the total number of signal events) as a function of $m_{\NI}$
for various beam energies $\sqrt{s} = 161$, 172, 185, 190, 195, and 200 GeV.
The efficiency decreases slightly with increasing beam energy, but always
exceeds 50\%. Also, we note that the efficiency increases slightly closer
to threshold; this is because nearer threshold the $M_{\rm INV}$ distribution
for the signal becomes somewhat more sharply peaked with a smaller
overlap with the cut region $M_{\rm INV} > 80$ GeV. The plotted
efficiencies were
found for a specific class of models with $\nmess=1$, $\tan\beta=3$, $x=0.1$,
$\mu < 0$ and varying $\Lambda$, but these results
are very nearly model-independent for a fixed $m_{\NI}$,
since the efficiencies
depend mostly on the kinematics of the isotropic $\NI$ decays. Therefore
one may estimate the usable cross section (after all cuts) for future
LEP2  runs
by simply multiplying the total cross section by 0.5.
While the LEP runs at $\sqrt{s} = 161, 172$
were limited by only having $\sim 10$ pb$^{-1}$ collected per experiment,
the reach at $\sqrt{s}=172$ GeV extends even beyond
$m_{\NI} = 77$ GeV in some parts of the GMSB model parameter space.
However, considering the minimum cross section obtained in some models
(see Fig.~\ref{figminmax}),
the exclusion capability for these runs is about
$m_{\NI} > 72$ GeV, using a criterion of 5 total events (summed over
all four detectors) after
all cuts. This limit is quite comparable to what should
be attainable with the present Tevatron data (and provides
{\it a posteriori}\/
justification for our assumption $m_{\NI} > 70$ GeV).
With 300 pb$^{-1}$ and a discovery requirement of 5 events after cuts,
the discovery reach should extend up to about $m_{\NI} = \sqrt{s}/2 - 4$
GeV in future runs, based on an efficiency of 50\%.
If a discovery is made, the events with the largest photon energies
can be used to find $m_{\NI}$ using eqs.~(\ref{photonenergyrange})
and (\ref{photonenergybounds}).
(The lower endpoint of the energy range will be contaminated with
$\gamma\gamma\nu\overline\nu$ background events.)

We note
that observation of a few $\gamma\gamma\Etot$ events
with $M_{\rm INV} > 80$ GeV could not
be interpreted as unambiguous evidence for $\NI\NI$ production,
since the background is comparable to or larger than the signal,
especially if our cut on soft photons (\ref{softphotoncut}) is not applied.
Conversely any events with $M_{\rm INV} < 80$ GeV would be of great interest
for the GMSB scenario,
since the physics backgrounds for such events are quite
negligible.\footnote{We also note that in the higgsino LSP interpretation
\cite{AKKMM1,AKKMM3}
of the CDF event \cite{Event}, one could conceivably have
$\gamma\gamma\Etot$ events
from $\NII\NII$ production with the one-loop decay
$\NII\rightarrow\gamma\NI$, but these would yield softer photons,
and like the background would tend to have larger $M_{\rm INV}$ than a
GMSB signal.}

When kinematically allowed, slepton pair production can add to the
main $\NI\NI$ discovery signal.
Each of the production cross-sections for $\ser\ser$, $\smur\smur$,
and $\stau_1\stau_1$ can even exceed $0.1$ pb at $\sqrt{s}=190$ GeV,
adding events $\ell^+\ell^-\gamma\gamma\Etot$ to the $\gamma\gamma\Etot$
signal we have just discussed. The leptons in these events should
be softer than the photons,
because the slepton production cross section cannot be significant
unless $m_{\slepton}-m_{\NI}$ is small. For some models with large
$\tan\beta$, only $\stau_1\stau_1$ pair production with
signal $\tau^+\tau^-\gamma\gamma\Etot$ can occur in addition to $\gamma\gamma
\Etot$. While the individual
slepton pair production cross sections never exceed
half of the $\NI\NI$ production cross section, the
$\ell^+\ell^-\gamma\gamma\Etot$ signal(s) should not have any significant
backgrounds (especially
considering that the photons are always energetic). Therefore
it should be kept in mind that
slepton pair production can be an important component of the discovery
signal even in the neutralino NLSP scenario.
In the minimal model with $n=1$, this can only
occur if $x$ is significantly greater than 0, so that observation of both
$\gamma\gamma\Etot$ and $\ell^+\ell^-\gamma\gamma\Etot$ at LEP2 would exclude
models with $n=1$ and small relative mass splittings for the messenger fields.

\subsection*{B. The slepton co-NLSP scenario} \indent

In this subsection we consider the case that $\stau_1$, $\ser$, and $\smur$ are
lighter than all of the other superpartners, and are nearly degenerate in mass.
As long as the conditions eq.~(\ref{codethree}) are satisfied, then $\stau_1$,
$\ser$, $\smur$ act effectively as co-NLSPs, each decaying directly to the
corresponding lepton plus gravitino.  
This situation can arise if $\nmess=2$,
3,
or 4, and $\tan\beta \lsim$ 8.  For larger values of $\tan\beta$, the tau
Yukawa
coupling causes mixing between the left- and right-handed staus which is always
sufficient to render $m_{\stau_1}$ lower than $m_{\smur}$ by more than the
$\tau$ mass.  Restricting our attention to models with slepton NLSP masses
between 50 and 100 GeV, we find that the allowed ranges for the parameter
$\Lambda$ are
\beq
& &\nmess=2;\qquad 15\>{\rm TeV}\> < \> \Lambda \> < \>42 \>
{\rm TeV}, \>\>{\rm or} \label{SLNLSP1}\\
& &\nmess=3; \qquad 11\>{\rm TeV}\> <
\> \Lambda \> < \>35 \> {\rm TeV},\>\>{\rm or} \label{SLNLSP2}\\
& &\nmess=4;\qquad 10\>{\rm TeV}\> < \> \Lambda \> < \>28 \> {\rm TeV}
\label{SLNLSP3}  \eeq
in order for eq.~(\ref{codethree}) to be satisfied.
As mentioned in the Introduction, however, there are some models for which
$m_{\slepton_R} - m_{\stau_1} - m_{\tau} - m_{\ell}$ can be up to a few GeV
but 
$\slepton_R \rightarrow\ell\G$ can still dominate if $\sqrt{F}$ is not
too large and $\slepton_R \rightarrow \ell\NI$ is not kinematically open. 
Those models will also
act as slepton co-NLSP models.
In any case, using eqs.~(\ref{sleptondecaylength}) and (\ref{Fexp}),
we find that the minimum possible decay length at LEP2  in the slepton
co-NLSP scenario is about 10 microns for $\nmess=2$, and somewhat smaller for
$\nmess=3,4$.  Of course, values of $x\lambda$ smaller than 1 will increase the
decay length proportionally to $1/(x\lambda)^2$, and $F_S<F$ would have the
same effect.  This means that the $\slepton \rightarrow \ell\G$ decay lengths
can easily exceed minimum detector resolutions, providing a
background-independent signal \cite{DDRT,DTW2} if the tracks of stable sleptons
and/or their macroscopic decay lengths are observed by the LEP2  detectors.
This would be spectacular confirmation of the GMSB scenario.

The cross section for $\smur\smur$ production at LEP2
(as a function of $m_{\smur}$) is
model-independent,
since this process is mediated only by $s$-channel exchange of $\gamma,Z$.
The $\ser\ser$ pair production cross section
has a contribution from $t$-channel
neutralino exchange, but there is significant destructive interference
with the $s$-channel $\gamma,Z$ exchange graphs, especially near threshold.
This means that $\sigma(\ser\ser)$ is often much lower than
$\sigma(\smur\smur)$ in slepton co-NLSP GMSB models. This is an
important qualitative difference from the situation in neutralino LSP
models as studied in \cite{CDMT,BBMT}, where exchange of
a lighter neutralino in the $t$-channel typically ensures that
$\sigma(\ser\ser) > \sigma(\smur\smur)$.
The destructive interference effect in slepton co-NLSP models is
greater for larger values of $n$, corresponding to heavier $\NI$.
We find that in our slepton co-NLSP model parameter space,
$\sigma(\ser\ser)<\sigma(\smur\smur)$ always holds for $m_{\slepton_R}$
more than about (10, 16, 20) GeV below the kinematic threshold of
$\sqrt{s}/2$, for models with $n=$ (2, 3, 4).

\begin{figure}[tb]
\centering
\epsfxsize=3.8in
\hspace*{0in}
\epsffile{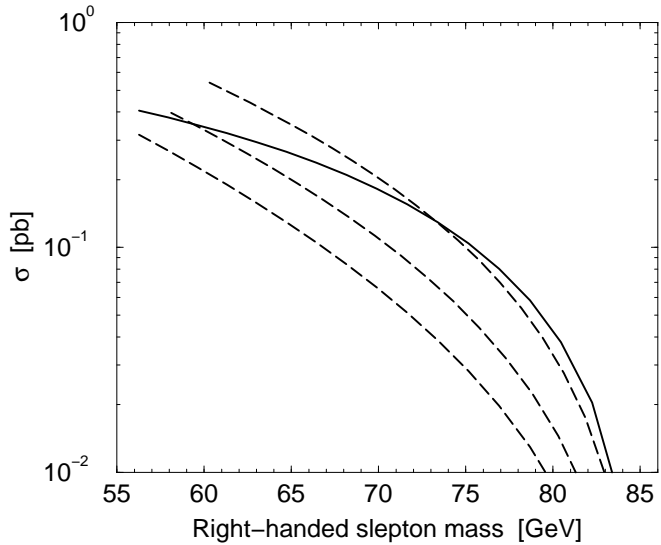}
\caption{Slepton pair production cross sections
as a function of slepton mass, for $e^+e^-$ collisions at
$\protect\sqrt{s} = 172$ GeV. The
solid curve is the (model-independent)
production cross section for $\smur\smur$.
The three dashed lines are the $\ser\ser$ production cross sections
for a family of models with varying $\Lambda$ and fixed
$x=0.1$, $\mu<0$, $\tan\beta=1.5$, and, from
top to bottom, $\nmess=2,3,4$. (The model dependence is due to the
$t$-channel exchange of neutralinos.) The $\stau_1\stau_1$ production
cross section in slepton co-NLSP
models is always nearly equal to that of $\smur\smur$.}
\label{figsleptonprod172}
\end{figure}
\begin{figure}[tb]
\centering
\epsfxsize=3.8in
\hspace*{0in}
\epsffile{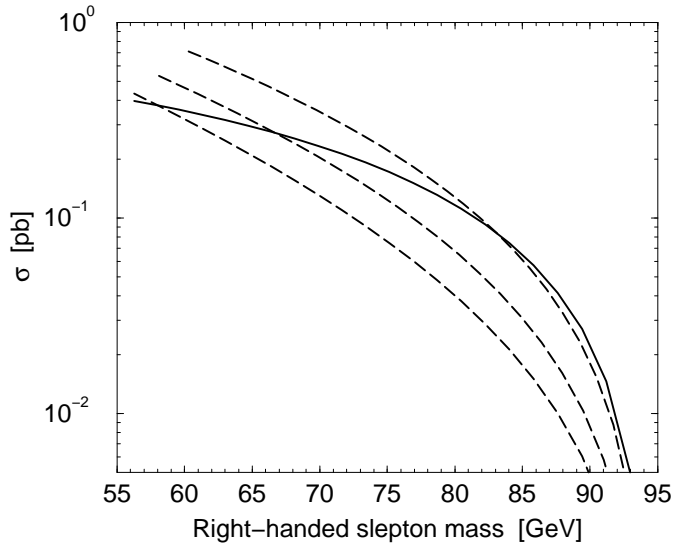}
\caption{As in Figure~\ref{figsleptonprod172}, but with $\sqrt{s}=190$
GeV.}
\label{figsleptonprod190}
\end{figure}
These features are illustrated in Figs.~\ref{figsleptonprod172} and
\ref{figsleptonprod190}, which show
production cross sections in
$e^+e^-$ collisions at $\sqrt{s} = 172$ and 190 GeV for
$\ser\ser$ and $\smur\smur$. The results shown for $\ser\ser$ are for
a typical family of GMSB models with
$x = 0.1$, $\tan\beta=1.5$, $\mu < 0$ and $\Lambda$ varying,
as a function of $m_{\smur} \approx m_{\ser}$.
The three dashed lines in Fig.~\ref{figsleptonprod172} and
Fig.~\ref{figsleptonprod190} are the $\ser\ser$
cross sections in this class of models for $\nmess=2,3,4$, from top to bottom.
(These models have lower bounds on $m_{\slepton_R}$ as indicated;
these follow
indirectly from a lower bound on the mass of the lightest Higgs boson,
which we take to be $m_h < \sin^2(\beta-\alpha)$ 64 GeV \cite{Higgsbound}.)
For a given $n$, the
$\ser\ser$ production cross sections in other models can be
up to a factor of two smaller, but not much  larger,
than shown in Figs.~\ref{figsleptonprod172}
and \ref{figsleptonprod190}.
The production cross section for $\stau_1\stau_1$ is
always nearly equal to
that for $\smur\smur$, owing to the small mixing required by low $\tan\beta$
and near-degeneracy of $\smur$ and $\stau_1$.
For a given model, one finds that $\sigma(\stau_1\stau_1)$ exceeds
$\sigma(\smur\smur)$ by a few per cent.

With the 10 pb$^{-1}$ collected at $\sqrt{s}=172$ GeV it is apparent
from Figure~\ref{figsleptonprod172} that,
modulo detector-dependent considerations regarding the rate of
energy loss
by ionization and tracking chamber capabilities, one should be able
to put a useful exclusion on long-lived sleptons.
Indeed, the DELPHI collaboration has recently analyzed data from
runs with $\sqrt{s}\leq 161, 172$ GeV, and found that
long-lived right-handed smuons and
staus with mass less than 65 GeV are excluded
at the 95\% confidence level \cite{DELPHIlimit}. A somewhat more restrictive
lower bound could presumably be obtained by combining the results of
all four detectors. Likewise, searches for sleptons with decay lengths
exceeding several centimeters could probably establish similar limits.
Comparing with Figure~\ref{figsleptonprod190}, we see that in future runs
with 300 pb$^{-1}$ or more, it should be possible to exclude or discover
long-lived sleptons with masses up to a few GeV of the kinematic limit.
(Comparable constraints on long-lived sleptons from the present Tevatron
data will probably be difficult to obtain \cite{Stuartpc}.)

\begin{figure}[tb]
\centering
\epsfxsize=3.8in
\hspace*{0in}
\epsffile{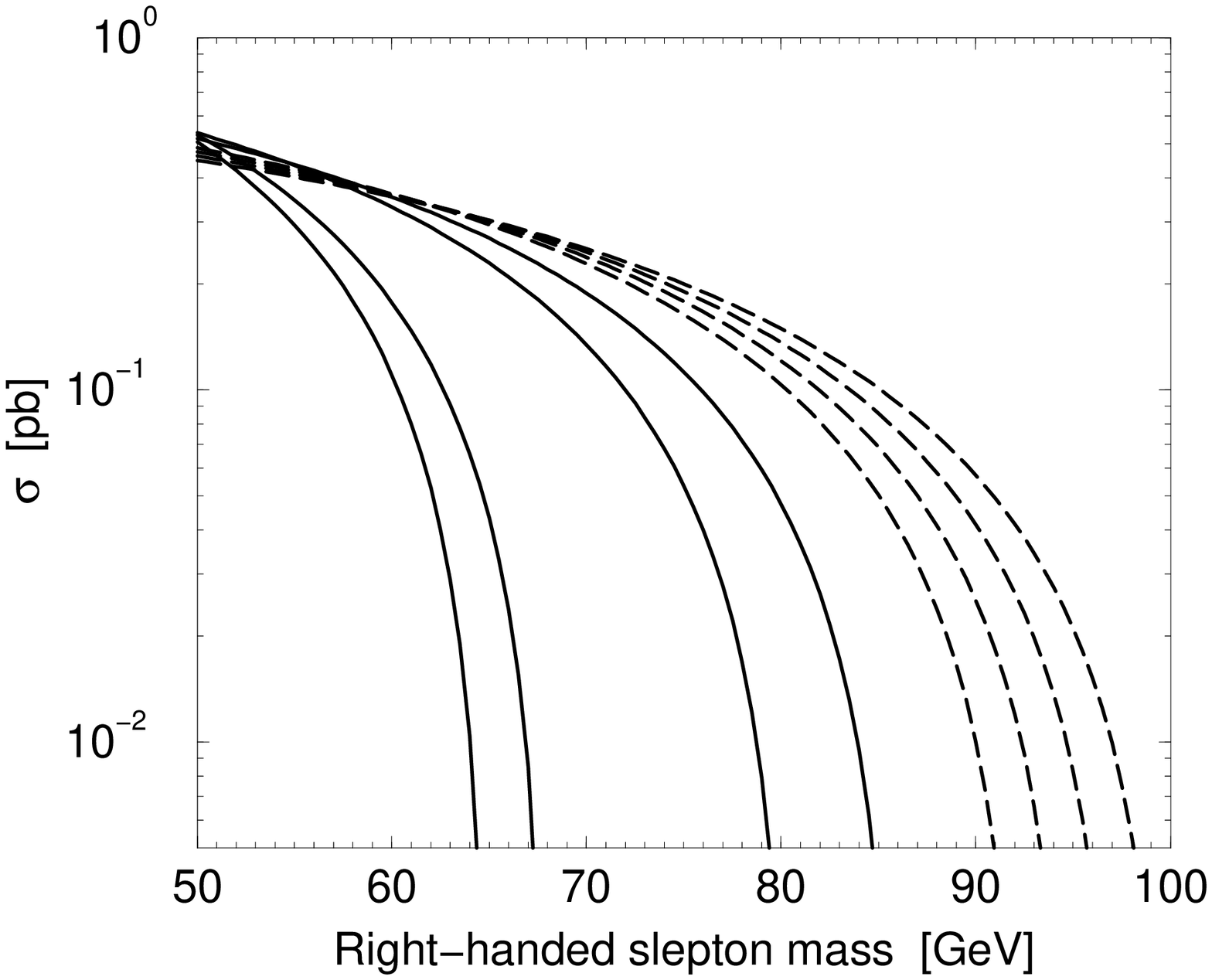}
\caption{Smuon pair production cross sections
as a function of $m_{\smur}$, for $e^+e^-$ collisions at
(from left to right)
$\protect\sqrt{s} = 130, 136, 161, 172$ (solid lines) and
185, 190, 195, and 200 GeV (dashed lines).}
\label{figsmuonprod}
\end{figure}

In the remainder of this subsection,
we concentrate on the more difficult situation that
the finite decay lengths for the sleptons are too short to measure.
The signals for
slepton pair production are then $e^+e^-\Etot$, $\mu^+\mu^-\Etot$, or
$\tau^+\tau^-\Etot$ for $\ser\ser$, $\smur\smur$, $\stau_1\stau_1$ respectively
\cite{DDRT}.
As illustrated by the examples of Figures \ref{figsleptonprod172} and
\ref{figsleptonprod190}, the worst-case situation for a given
$m_{\slepton_R}$ will have $\sigma(\ser\ser) \ll \sigma(\smur\smur)$,
so we concentrate on $\smur\smur$ as the discovery process. (In general,
$\ser\ser$ can be the dominant discovery process
in slepton co-NLSP models only when both cross sections are
large and discovery is relatively easy anyway.)
In Figure~\ref{figsmuonprod},
we show the model-independent production cross section for $\smur\smur$
as a function of
$m_{\smur}$, for various beam energies.
The past LEP runs at $\sqrt{s}=130$-136, 161
and 172 GeV collected about 5.7 pb$^{-1}$, 10 pb$^{-1}$ and 10 pb$^{-1}$
per detector respectively. Slepton masses less than about 55 GeV would
have resulted in several $\ell^+\ell^-\Etot$ events for each detector
at LEP130-136, with a detection efficiency probably well in excess of 50\%,
since the
decays $\slepton\rightarrow\ell\G$ will always result in energetic leptons.
(See for example the analogous situation analyzed in \cite{ALEPHOPAL}
in the case of sleptons decaying to a lepton and light neutralino.)
For the LEP161 and LEP172 runs, several events per detector could
be expected for slepton masses up to perhaps 70 GeV, but with a background from
$W^+W^-$ production with leptonic $W$ decays, as discussed below. A precise
determination of slepton mass exclusions from present data
in the slepton co-NLSP case would
involve detector-specific issues and will not be attempted here.

Future runs with higher
beam energy and much more data should be able to decisively probe
a significant range of slepton masses.
The lepton energies in each event have a flat distribution (before any cuts)
with endpoints given by
\beq
E_{\rm min} < E_{\ell^+},E_{\ell^-} < E_{\rm max}
\label{leptonenergyrange}
\eeq
where
\beq
E_{\rm max, min} = {1\over 4}({\sqrt{s}} \pm \sqrt{s - 4 m_{\slepton}^2})\> .
\label{leptonenergybounds}
\eeq
Therefore the leptons from the signal events are quite energetic
(especially in the critical case that
$m_{\tilde\ell}$ is near the kinematic threshold so that the production
cross section is low),
allowing one to choose a rather strong cut on the minimum lepton energy.
To reduce an important component of the background as discussed below,
it is necessary to impose such a lower bound on lepton energy (in contrast
to the situation for slepton signals with a neutralino LSP \cite{CDMT,BBMT},
where it is instead useful to impose an upper bound cut on lepton energies).
We will somewhat arbitrarily take this to be
\beq
E_\ell > 20\>\>{\rm GeV},
\label{leptonenergycut}
\eeq
which does not impact the signal at all for $m_{\slepton} > 77.5$ GeV for
$\sqrt{s} = 190$ GeV,
but this can and should
be adjusted depending on the signal and on the collider parameters.
In the following discussion we also impose a detectability cut
\beq
|\eta_\ell| < 2.5
\label{leptonanglecut}
\eeq
on the pseudorapidity of each
lepton.

The fraction of $e^+e^-\Etot$, $\mu^+\mu^-\Etot$ and $\tau^+\tau^-\Etot$
signal events which pass these cuts is always quite high.
Unfortunately, there are significant
backgrounds to these signals which must be considered.
First, one has
$e^+ e^- \rightarrow \tau^+\tau^-$ with
leptonic $\tau$ decays which gives a background to $\ell^+\ell^-\Etot$
with $\ell=e,\mu$.
The resulting leptons are always nearly
back-to-back, so that an acoplanarity cut
\beq
\cos\phi(\ell^+\ell^-) > -0.9
\label{acoplanaritycut}
\eeq
can essentially eliminate this background, while leaving the signal
largely intact. (The angle $\phi$ is defined between the two leptons in
the plane transverse to the beam axis.)
Backgrounds from $\ell^+\ell^-\gamma$ and $\ell^+\ell^-e^+e^-$ production with
the photon or $e^+e^-$ lost down the beampipe can be efficiently
eliminated   with a cut
on the total missing transverse momentum \cite{CDMT}:
\beq
\slashchar{p}_T > 0.05 \sqrt{s} .
\label{missingtransversecut}
\eeq
Significant backgrounds also arise from $e^+e^-\rightarrow
Z(\rightarrow\nu\overline\nu)
\gamma^*(\rightarrow\ell^+\ell^-)$ and  $e^+e^-\rightarrow$ $
Z(\rightarrow\nu\overline\nu)Z(\rightarrow\ell^+\ell^-)$.
To eliminate them we impose a cut
\beq
|M_{\rm INV} - M_Z| > 10\>\>{\rm GeV}
\label{minvsleptoncut}
\eeq
on the missing invariant mass $M_{\rm INV} \equiv (p_{e_i^+} + p_{e_i^-} -
p_{\ell_f^+} - p_{\ell_f^-})^2$ in each event.
Other minor
backgrounds are present from higher order processes such as
$e^+e^-\rightarrow e\nu W(\rightarrow e\nu)$. Their sizes before and after
our cuts are given in Table 2, where we show that they are reduced to
a negligible level. It should be noted that these cuts will also reduce
any potentially dangerous interferences of these backgrounds with
the main background we are about to discuss.

When $\sqrt{s} > 2 M_W$, the largest physics background is due to
$W$-pair production followed by leptonic $W$ decays:
\beq
e^+e^- \rightarrow W^+ W^- \rightarrow \ell^+\ell^-\nu\overline\nu .
\label{directWWbackground}
\eeq
At $\sqrt{s} = 190$ fb, this background amounts to 238 fb for each lepton
flavor, before cuts. This is reduced to 140 fb after applying the cuts
eqs.~(\ref{leptonenergycut})-(\ref{minvsleptoncut}).
The kinematics of the background (\ref{directWWbackground})
are similar to those of the slepton
pair-production signal, especially
if $m_{\slepton}$ is close to $m_W$, and in
particular the cut (\ref{leptonenergycut})
has only a very small effect.
The situation is rather similar to the case of slepton pair production
at LEP in the neutralino LSP scenario as studied in \cite{CDMT,BBMT}, but with
$\slepton\rightarrow\ell\G$ taking the place of $\slepton\rightarrow\ell\NI$.
Note that in the present situation,
the near masslessness of the gravitino makes the decay
$\slepton\rightarrow\ell\G$ even more kinematically similar to the
Standard Model decay
$W\rightarrow\ell\nu$.  In particular, the signal
cannot be enhanced significantly by imposing an upper bound
on the lepton energies, as it could be
in the situation investigated in \cite{CDMT,BBMT}.
However, we can still use the fact that the positively (negatively)
charged leptons from $W^+W^-$ production are produced preferentially
in the same direction as the positron
(electron) beam.
This polar angle
asymmetry also unfortunately holds true for the $e^+e^-\Etot$ signal
from $\ser\ser$ production, although not as strongly as for the background.
The $\mu^+\mu^-\Etot$ and $\tau^+\tau^-\Etot$
signals are fortunately
symmetric with respect to $\theta \rightarrow \pi-\theta$
because $e^+e^-\rightarrow\smur\smur$ and
$e^+e^-\rightarrow\stau_1\stau_1$ production
have only $s$-channel contributions.
Therefore the signal/background ratio for $\smur\smur$
and $\stau_1\stau_1$ (and to a lesser extent $\ser\ser$)
can be significantly
enhanced by imposing a cut
\beq
\pm \cos\theta_{\ell^\pm} > 0
\label{asymmetrycut}
\eeq
on the more energetic lepton in each event, as in \cite{CDMT,BBMT}. (We always
use the definition of $\theta$
as the angle between the $e^-$ beam momentum and the outgoing lepton
momentum.)

We must also consider
a background contribution for
$e^+e^-\Etot$ and $\mu^+\mu^-\Etot$ (but not for $\tau^+\tau^-\Etot$)
from
\beq
e^+e^-\rightarrow W^+W^- \rightarrow \ell^\pm\tau^\mp
\nu\overline\nu \rightarrow \ell^+ \ell^- \nu\overline\nu\nu\overline\nu
\label{indirectWWbackground}
\eeq
Since $\tau\rightarrow\ell\nu\overline\nu$ has a branching fraction
of about 0.18 for each of $\ell=e,\mu$,
a rough estimate is that the additional $\ell^+\ell^-\Etot$
backgrounds are each about 0.36 times the direct
$W^+W^-\rightarrow\ell^+\ell^-\nu\overline\nu$ background.
However, this is an overestimate, since
the resulting leptons tend to be softer so that
a large fraction of these events are
eliminated by the minimum energy cut
(\ref{leptonenergycut}), which was included for this reason.
Before (after) the cuts eqs.~(\ref{leptonenergycut})-(\ref{minvsleptoncut})
this background contributes 84 (18) fb.
It should be noted that if some efficient tau tagging is possible, then some
part of this background could be eliminated. However, this is a highly
detector-dependent matter and we choose to simply consider the whole
background.
There is also a background from
$e^+e^-\rightarrow W^+W^-\rightarrow\tau^+\tau^-\nu\overline\nu \rightarrow
\ell^+\ell^-\nu\overline\nu\nu\overline\nu$, but this is suppressed by
the factor $BR(\tau\rightarrow\ell\nu\overline\nu)^2 \approx$ 0.03 and
is greatly diminished further
by the cut (\ref{leptonenergycut}), and so can be safely neglected.
In Table~2 we summarize the dominant $\ell^+\ell^-\E$ backgrounds showing the
effects of the cuts described above.

To evaluate the background from eq.~(\ref{indirectWWbackground}) we used
our own Monte Carlo simulating the tau decay following each of the parent
$WW\rightarrow\mu\tau\nu\nu$ events
generated using CompHEP and the BASES/SPRING package \cite{BasesSpring}.
In doing this, we also took into account
spin correlations and tau-decay
anisotropies in the tau rest frame.
We checked that neglecting the latter effects
would have resulted in an 18\% underestimate of this
background after cuts (1-5) in Table 2 and in an 
only slightly flatter
distribution
in $\pm\cos\theta_{\ell^\pm}$ so that the underestimate is diminished to 17\%
after cuts (1-6). The effect of taking into account spin correlations
is not dramatic in this case because the
pattern of distributions is dominated by a large boost of the tau in the
lab frame.

\begin{figure}[tb]
\centering
\epsfxsize=3.7in
\hspace*{0in}
\epsffile{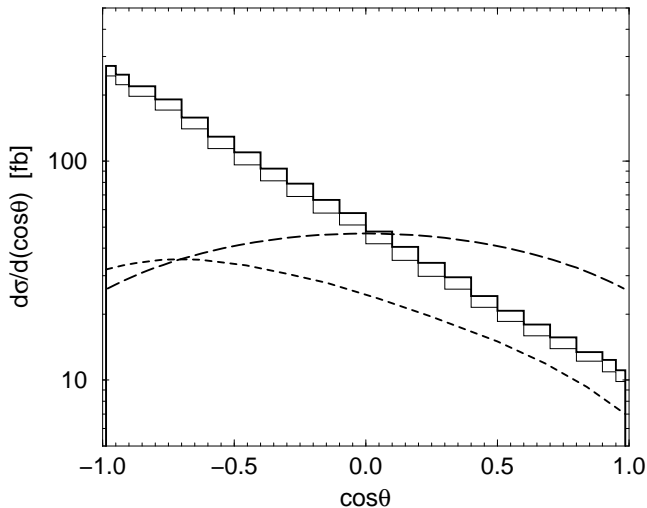}
\caption{The differential cross section $d\sigma/d(\pm\cos\theta_{\ell^\pm})$
for the more energetic
lepton in each event at
$\protect\sqrt{s} = 190$ GeV, after the cuts
(\ref{leptonenergycut})-(\ref{minvsleptoncut}).
The long-dashed (short-dashed)
line is the signal from $e^+e^- \rightarrow \tilde\ell^+ \tilde\ell^-
\rightarrow \ell^+\ell^-\Etot$ for muons (electrons), respectively.
For this figure we have chosen a model with $m_{\slepton_R} = 80$ GeV
as described in the text.
The lighter solid line histogram shows
the background contribution from $e^+e^- \rightarrow
W^+W^- \rightarrow \ell^+\ell^-\nu\overline\nu$ (not summed
over lepton flavors), while the heavier solid line histogram includes also
the background from $e^+e^- \rightarrow W^+W^- \rightarrow \ell^\pm\tau^\mp
\nu\overline\nu \rightarrow \ell^+ \ell^- \nu\overline\nu\nu\overline\nu$
where $\ell=e$ or $\mu$.}
\label{pmcosthetafigure}
\end{figure}

These considerations are illustrated in Figure \ref{pmcosthetafigure},
which shows the distribution
of $\pm \cos\theta_{\ell^\pm}$
for the $\smur\smur$ and $\ser\ser$ signals in a model
with $m_{\slepton_R} = 80$ GeV. (The other model parameters, which
do not affect the distribution from smuons, were chosen to be
$\nmess=3$, $\Lambda = 24$ TeV, $\tan\beta=1.5$, $x=0.1$,
with $\mu=-460$ GeV.)
Also shown as the heavier solid line histogram
is the distribution for background events from
both (\ref{directWWbackground}) and (\ref{indirectWWbackground}).
The $\mu^+\mu^-\Etot$ component of the signal is much more promising
than $e^+e^-\Etot$,
both because the total cross section is larger
and because the polar angular distribution is less similar to the background.
This is a quite general feature.
In the case of the $\tau^+\tau^-\Etot$
signal, there is no background from (\ref{indirectWWbackground}), so we
also show in Figure \ref{pmcosthetafigure} the distribution from
(\ref{directWWbackground}) only, as the lighter solid line histogram.
(The $\tau^+\tau^-\Etot$ signal is, however, subject to a significant
and quite detector-dependent loss from $\tau$ identification efficiencies.)

\begin{figure}[tb]
\centering
\epsfxsize=3.7in
\hspace*{0in}
\epsffile{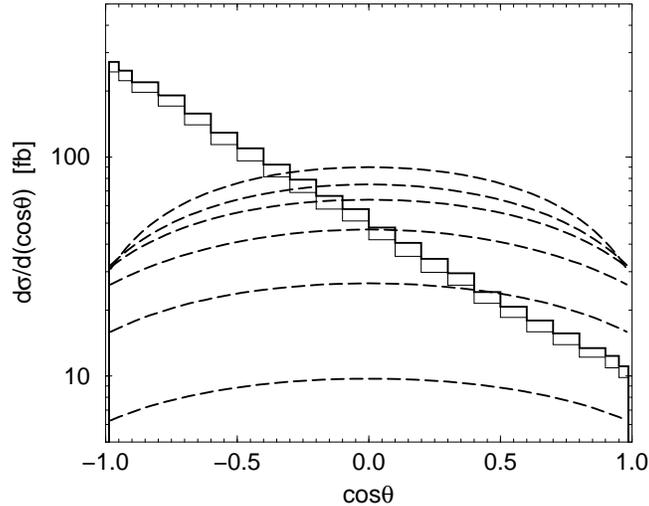}
\caption{The differential cross section $d\sigma/d(\pm\cos\theta_{\ell^\pm})$
for the more energetic muon in each event at $\protect\sqrt{s} = 190$ GeV,
after the cuts (\ref{leptonenergycut})-(\ref{minvsleptoncut}).
The dashed
lines are the signal from $e^+e^- \rightarrow \smur\smur \rightarrow
\mu^+\mu^-\Etot$ for, from the top down, $m_{\smur} =$ 60,
70, 75, 80, 85, 90 GeV.
The background distributions are shown as described in Figure
\ref{pmcosthetafigure}.}
\label{pmcosthetafiguresmuon}
\end{figure}

As these examples illustrate,
we may concentrate on $\smur\smur$
production as the likely discovery process, at least
in the pessimistic but
common situation that $\sigma(\ser\ser)\lsim \sigma(\smur\smur)$.
In Figure \ref{pmcosthetafiguresmuon} we compare the distributions
for the polar angle of the more energetic muon from background and
signal events, for various $\smur$ masses.
The $W^+W^-$ background should contain $e^+e^-\Etot$, $\mu^+\mu^-\Etot$,
$e^+\mu^-\Etot$  and $e^-\mu^+\Etot$ events in equal amounts.
Meanwhile the signal yields more
$\mu^+\mu^- \Etot$ than $e^+e^-\Etot$ events after the cut
eq.~(\ref{asymmetrycut}), and
no $e\mu\Etot$ events [except from $\stau_1\stau_1$ production
with both taus decaying leptonically; this signal is again
strongly diminished by the square of the
leptonic branching fraction of the tau
and by the energy cut eq.~(\ref{leptonenergycut})].
The cross sections at $\sqrt{s} = 190$ GeV for signal $\mu^+\mu^-\Etot$
events before cuts are (360, 238, 178, 121, 69, 25) fb for
$m_{\smur} =$ (60, 70, 75, 80, 85, 90) GeV, respectively.
The corresponding amounts
which pass all of the cuts including eq.~(\ref{asymmetrycut}) are
(70, 60, 52, 39, 23, 8.5) fb respectively.
This amounts
to an expectation of (21, 18, 16, 12, 7, 2.5) signal
$\mu^+\mu^-\Etot$ events per detector with 300 pb$^{-1}$.
For comparison, the total $\mu^+\mu^-\Etot$
background from eqs.~(\ref{directWWbackground}), (\ref{indirectWWbackground})
and other sources
which survives all of the
cuts including eq.~(\ref{asymmetrycut})
is 26 fb (see Table 2),
or 7.8 events in 300 pb$^{-1}$.
Combining the results of all four detector should therefore give a 5$\sigma$
discovery signal for $m_{\slepton_R}$ up to 85 GeV.
However, we conclude from
Fig.~\ref{pmcosthetafiguresmuon} that
an unambiguous discovery may not be possible for models with
$m_{\smur} \gsim 85$ GeV,
even with 300 pb$^{-1}$ per experiment at  $\sqrt{s} =$ 190 GeV,
because of the low rates and significant $WW$
background.\footnote{In the above analysis we
did not take into account initial
state radiation effects, which
diminish total cross sections for both the signal and the backgrounds
by up to 10\%, and produce slight changes in the shapes of distributions.
We do not expect this to significantly affect our conclusions.}

In the delicate region where one is searching for smuons with
$m_{\smur} > 80$ GeV, it will help to increase the minimum
muon energy cut eq.~(\ref{leptonenergycut}) and to impose an upper
bound cut on muon energies, in view of
eq.~(\ref{leptonenergybounds}), since this will reduce
the $W^+W^-$ backgrounds somewhat. It may also be useful
to adjust the
polar angle cut (\ref{asymmetrycut}).
The optimal cuts clearly depend in a non-trivial way on the beam energy,
on the amount of integrated luminosity
available, and on the masses of the sleptons being searched
for, because of the low signal rates.
If a discovery is made, the events with the highest muon energy
eq.~(\ref{leptonenergycut}) should provide the best
estimate of the smuon mass using (\ref{leptonenergybounds}),
if $m_{\smur} < 80$ GeV.
For $m_{\smur} \gsim 80$ GeV, such a
determination will be more difficult because of the lower rates and
because the range of muon energies from the signal events is entirely
covered by the background.

It is important to note that when kinematically allowed, $\NI\NI$ will
likely provide the clearest discovery signal. Despite $m_{\NI} > m_{\stau_1},
m_{\ser}, m_{\smur}$ in these models, we find that $\sigma(\NI\NI)$ can
be as much as 3 times larger than $\sigma(\smur\smur)$ in the
slepton co-NLSP scenario. Each $\NI$ can decay
as $\NI \rightarrow \slepton_R\ell \rightarrow \ell^+\ell^-\G$
or $\NI \rightarrow \stau_1 \tau \rightarrow \tau^+\tau^-\G$.
The first lepton
emitted in the decay chain will often be very soft
(if $m_{\NI} - m_{\slepton_R}$ or $m_{\NI} - m_{\stau_1} - m_{\tau}$
is small), but this does not
degrade the signal. The reason is that because of the Majorana nature
of $\NI$, the charges and flavors
of the two most energetic leptons in the event
(which carry most of the visible energy) are uncorrelated with each other.
Therefore one has the rather unique signature of
$\ell^+\ell^{\prime+} (\ell^-\ell^{\prime-}) \Etot$ and
$\ell^-\ell^{\prime-} (\ell^+\ell^{\prime+}) \Etot$ and
$\ell^+\ell^{\prime-} (\ell^-\ell^{\prime+}) \Etot$ events
in the ratio 1:1:2,
with the leptons in parentheses being much softer, perhaps even too soft
to detect. Here $\ell$ and $\ell^\prime$ can each be $e$, $\mu$, or $\tau$,
independently.
Even if both soft leptons are not detected, the presence of two energetic
same-charge
leptons with large missing energy should be an unmistakable discovery
signal for the first two of these signatures.
(The analog of this signal in the special case of
a stau NLSP scenario was recently discussed in \cite{DDN}.)
The $\NI\NI$ production cross section in slepton co-NLSP models is still
bounded from below as in Figure \ref{figminmax} (as a function of
$m_{\NI}$), but can also be much larger. We therefore find that
the $\NI\NI$ channel will generally provide the largest discovery signal
in slepton co-NLSP models
if $m_{\NI}$ is more than about 5 GeV below threshold.
This is particularly likely to happen in models with $n=2$, for which
$\NI$ tends to be not much heavier than the sleptons.
It is also especially probable
when $\sigma(\ser\ser)$ and $\sigma(\smur\smur)$
are also both large, so that identifying
an excess of $e^+e^-\Etot$ and
$\mu^+\mu^-\Etot$ events above the $WW$ backgrounds will also not be difficult.
In that case the slepton pair production and $\NI\NI$ pair production
signals should provide strong confirmation of each other.

\subsection*{C. The stau NLSP scenario}
\indent

In this section we consider the case that
the lightest stau is the NLSP
and all supersymmetric decay chains terminate in $\stau_1\rightarrow
\tau\G$.
Restricting our attention to $\stau_1$ masses
between 50 and 100 GeV, we find that the condition eq.~(\ref{codetwo})
can be satisfied in the GMSB model parameter space of section II only
if
\beq
& &\nmess=1;
\qquad
35\>{\rm TeV}\> < \>\Lambda<\> 120\>{\rm TeV};
\qquad\tan\beta > 18,\>\>{\rm or}
\\
& &\nmess=2;
\qquad
18\>{\rm TeV}\> < \>\Lambda<\> 80\>{\rm TeV};
\qquad\tan\beta > 6,\>\>{\rm or}
\\
& &\nmess=3;
\qquad
12\>{\rm TeV}\> < \>\Lambda<\> 70\>{\rm TeV};
\qquad\tan\beta > 5,
\>\> {\rm or} \\
& &\nmess=4;
\qquad
10\>{\rm TeV}\> < \>\Lambda<\> 60\>{\rm TeV};
\qquad\tan\beta > 4.
\eeq
These requirements are necessary but not sufficient for the stau NLSP
scenario; indeed, as we have already mentioned,
$\slepton_R\rightarrow \ell\G$ for $\ell=e,\mu$ can dominate even if
eq.~(\ref{codetwo}) holds, provided that $\sqrt{F}$ is not
too large and $\slepton_R \rightarrow \ell\NI$ is not open.

As in the slepton co-NLSP scenario of the previous subsection, the discovery
prospects are clearest if the decay length $\stau_1\rightarrow \tau\G$
is macroscopic, so that the discovery signal  consists of
tracks from a heavy $\stau_1$ and/or kinks
due to $\stau_1$ decays which can be directly observed in
the detector \cite{DDRT,DTW2}.
In this case, there are no significant physics backgrounds,
and the discovery potential is limited only by the total production
cross section, the integrated luminosity, and the capabilities
of the detectors.
In the GMSB models with a stau NLSP,
the cross section for $e^+e^-\rightarrow \stau_1\stau_1$ as a function of
$m_{\stau_1}$ is given to a good approximation by the curves in Figure
\ref{figsmuonprod} with the horizontal axis now interpreted as $m_{\stau_1}$,
and in particular is not very model-dependent.
The mixing in the stau (mass)$^2$ matrix does provide for a small
reduction in the $\stau_1\stau_1$
cross section compared to that shown, but we have checked that this
reduction is at most about 10\% in our models.
As can therefore
be seen from Figure \ref{figsmuonprod}, future LEP2 runs should
be able to discover a long-lived stau NLSP with mass
up to close to the
kinematic limit, given 300 pb$^{-1}$ or more.

In the following, we therefore concentrate on the
more difficult possibility
that the $\stau_1$ decay length is too small to be directly
observed.
All events will then have an energetic
$\tau^+\tau^-$ pair and large missing energy.
In the large $\tan\beta$ limit,
one has $m_{\stau_1} \ll m_{\ser}, m_{\smur}, m_{\NI}$
in these models, so that the only discovery signal at
LEP2  is $\tau^+\tau^-\Etot$.
This must be compared to a background from
$W^+W^-$ with $W\rightarrow\tau\nu$ decays, as given in the previous section.
Since the signal events feature $E_\tau$ with a
flat distribution as in eqs.~(\ref{leptonenergyrange}) and
 (\ref{leptonenergybounds}),
the tau decay length of
$\sim 90\>\mu$m $\times E_\tau/m_\tau$  (roughly 1 to 4 millimeters)
may allow for fairly efficient tagging of non-leptonic $\tau$ decays.
Of course, the $W^+W^-$ background produces taus with a similar energy
distribution.
Just as discussed in the previous section, the
$\stau_1\stau_1$ signal is symmetric with respect to $\theta\rightarrow
\pi-\theta$, while the background is not.
Modulo the tau identification problem, Figure \ref{pmcosthetafiguresmuon}
gives an indication of the polar angle
distributions of signal and background for
$\tau^+\tau^-\Etot$ events. Extraction of a $\stau_1\stau_1$ signal from
the background will be considerably more difficult than would be the case for
$\smur\smur$ signal in the slepton co-NLSP scenario, and an estimate
of the reach will depend quite sensitively on detector capabilities.
If the taus are not tagged, one possibility is to look for purely hadronic
states with very large missing energy to avoid $WW$ contamination.

For lower $\tan\beta$, the production of $\ser\ser$, $\smur\smur$, and/or
$\NI\NI$ can also be kinematically allowed.
For a given model,
the $\smur\smur$ cross section is always smaller than that for
$\stau_1\stau_1$
production, due simply to the kinematic
suppression associated with $m_{\smur} > m_{\stau_1}+m_\tau$.
It
can of course be read off of Figure \ref{figsmuonprod}, as before.
The $\ser\ser$ production cross section can be either smaller or larger,
due to the interference between graphs
with $t$-channel exchange of neutralinos and $s$-channel exchange of
$\gamma,Z$. If  $\tan\beta$ does not exceed
about 30, these cross sections can add to the signal.
Now because of the decays $\slepton_R\rightarrow\ell\tau\stau_1$,
the signal from $\slepton_R\slepton_R$ production is
$\tau^+\tau^- (\ell^+\ell^{-}\tau^+\tau^-) \Etot$, where the leptons
in parentheses are much softer. If the leptons are energetic enough to be
identified, this signal should have very low backgrounds. It is also
possible that both $\slepton_R \rightarrow \stau_1 \tau\ell$ and
$\slepton_R \rightarrow \ell\G$ have a significant branching fraction,
if the former decay is sufficiently kinematically suppressed and
$\slepton_R \rightarrow \ell\NI$ is not open. The $\slepton_R\slepton_R$
production could then lead to the additional signatures
$\ell^+\tau^+(\ell^-\tau^-)\Etot$,
$\ell^-\tau^-(\ell^+\tau^+)\Etot$,
$\ell^+\tau^-(\ell^-\tau^+)\Etot$,
$\ell^-\tau^+(\ell^+\tau^-)\Etot$,
and, as in the slepton co-NLSP scenario, $\ell^+\ell^-\Etot$.

If $\NI\NI$ production is allowed it can
provide the dominant signal. In this case,
each $\NI$ can decay 
predominantly to $\tau\stau_1$ and then to $\tau^+\tau^-\G$.
The final taus from each $\stau_1$ decay will combine to carry most of
the visible energy in each event, and their charges are uncorrelated because
of the Majorana nature of $\NI$. This provides for the striking signatures
\cite{DDN}
$\tau^+\tau^+(\tau^-\tau^-)\Etot$ and
$\tau^-\tau^-(\tau^+\tau^+)\Etot$ and
$\tau^+\tau^-(\tau^+\tau^-)\Etot$ in the ratio 1:1:2, with the parentheses
denoting soft particles as before.  As in the case of slepton co-NLSP models,
we find that the $\NI\NI$ cross section can be up to 3 times larger than
that of $\stau_1\stau_1$, despite the requirement
$m_{\NI} > m_{\stau_1}+m_\tau$.
When $\tan\beta$ is not too large, this signal can be the most visible
one at LEP2 for stau NLSP models.
If the decays $\NI\rightarrow\ell\slepton_R$ are also allowed for
$\ell=e,\mu$, one may obtain the same signatures but with two or
four additional soft
leptons from the cascade decays of $\NI$ through $\slepton_R$. 
In our stau NLSP models we find that the $\NI\NI$ production cross
section as a function of $m_{\NI}$ is still bounded from below
(but can be up to 50\% smaller than indicated
in Figure \ref{figminmax} in some cases).
If $m_{\NI} \lsim \sqrt{s}/2 - 5$ GeV, there should be at least
a few events with very energetic same-charge taus and a pair of softer
taus with the opposite charge, if more than 300 pb$^{-1}$ is obtained.
This is again especially likely to be the
discovery signal in models with smaller $n$ and $\tan\beta$ not too large.

\subsection*{D. The neutralino-stau co-NLSP scenario}
\indent

Finally we consider the case that the lightest neutralino
and the lighter stau act effectively as co-NLSPs. This scenario will
occur if
the conditions of eq.~(\ref{codefour}) are satisfied.
While it might seem at first that requiring $m_{\NI}$ and $m_{\stau_1}$
to be nearly degenerate requires some fine-tuning, we find that the region
of parameter space where this occurs is actually
quite significant. Conditions
which are necessary for a viable model (with $m_{\NI} \gsim 70$ GeV)
in the neutralino-stau co-NLSP scenario are:
\beq
& &n=1; \qquad 40 \>{\rm TeV}\> < \> \Lambda < \> 80\> {\rm TeV};\qquad
15 < \tan\beta < 40, \>\>{\rm or} \label{tNLSP1} \\
& &n=2; \qquad 20 \>{\rm TeV}\> < \> \Lambda < \> 40\> {\rm TeV};\qquad
\tan\beta < 20, \>\>{\rm or} \label{tNLSP2} \\
& &n=3; \qquad 17 \>{\rm TeV}\> < \> \Lambda < \> 25\> {\rm TeV};\qquad
\tan\beta < 12 . \label{tNLSP3}
\eeq

Because $\NI$ and $\stau_1$ masses are not very different in this scenario,
strict exclusion or discovery should be rather straightforward for
a given NLSP mass. This is easily understood in terms of the
other scenarios we have already studied.
We find that in neutralino-stau co-NLSP models, the photino
content of $\NI$ is bounded from below by $\kappa_{1\gamma} > 0.35$.
Comparing eqs.~(\ref{neutralinodecaylength}) and
(\ref{sleptondecaylength}) we therefore conclude
that if the $\NI$ decay always
occurs outside of the detector, then the $\stau_1$ decay {\it must} also
take place over a typically macroscopic distance.
If $\NI\NI$ production is kinematically allowed,
then as in section III.A one has a signal $\gamma\gamma\Etot$
with about a 50\% detection efficiency after cuts,
as long as the $\NI$ decay length is not too long.
Conversely, if the decay lengths are long, then $\stau_1\stau_1$
production leads to a background-free signal with tracks of
quasi-stable $\stau_1$ observed directly in the detector using
the rate of energy loss and/or kinks due to slow decays $\stau_1\rightarrow
\tau\G$.
In an intermediate regime, one should see both types of events.
Note that in any case one need not rely on identifying a
$\tau^+\tau^-\Etot$ (or $\ell^+\ell^-\Etot$) signal against the
$W^+W^-$ background to effect discovery.
In neutralino-stau co-NLSP models, we find that the $\NI\NI$ production
cross sections at LEP2 are bounded from below as in Figure~\ref{figminmax}.
Likewise Figure \ref{figsmuonprod}
can be used to estimate the $\stau_1\stau_1$ production
cross section as a function of $m_{\stau_1}$. This overestimates
the actual $\stau_1\stau_1$ rate by no more than 10\% because of stau
mixing effects.
When $m_{\stau_1}$ and $m_{\NI}$ are
more than about 5 GeV below threshold, it cannot be possible
for both of these signals to elude detection in future
LEP2 runs.

\section*{IV. Concluding remarks}
\indent

In this paper we have studied the implications of gauge-mediated
supersymmetry breaking models for LEP2 physics. There are four
main scenarios for the effective NLSP(s),
each with its own predictions
for the possible discovery signals. These possibilities also depend on the
unknown NLSP decay length, and are summarized in Table~3.

In many cases, strict exclusion (or discovery!) for a given NLSP mass
can be assured given a sufficient beam energy and integrated luminosity.
For example, we found that $\sigma(\NI\NI)$ is bounded
from below (for a given $m_{\NI}$)
in the neutralino NLSP scenario models discussed in Section II.
We analyzed the backgrounds and found that they can be eliminated with
cuts which retain at least 50\% of the signal events.
This should guarantee discovery of $\NI$ with masses up to a few GeV of
the kinematic limit in future LEP2 runs, provided that the $\NI$ decay
length is not too long.
As a caveat, we must note that with variations of the model choices
we have made, it is quite possible to find smaller cross sections for
$\NI\NI$ production. For example, there could be
additional corrections to the Higgs soft (mass)$^2$ parameters
with an indirect result of smaller values for $|\mu|$ \cite{DTW2}.
Similarly, we have investigated some models with
unequal non-zero
values for the messenger multiplicities
$\sum_in_1(i)$, $\sum_in_2(i)$ and $\sum_in_3(i)$
[instead of eq.~(\ref{defnmess})],
and found that the requirements of correct electroweak symmetry breaking
can and do
lead to smaller values of $|\mu|/M_2$ in viable neutralino NLSP models.
In both cases, one finds that
$\NI$ can have a large higgsino content,
with therefore an arbitrarily small
production cross section for a given $m_{\NI}$
and a very long decay length for ${\NI} \rightarrow\gamma\G$.

In the slepton co-NLSP scenario, we found that if decays are prompt,
the $\mu^+\mu^-\Etot$ signal from $\smur\smur$ production is likely to
be the discovery mode. In particular, we found that the $e^+e^-\Etot$
signal from $\ser\ser$ production can be highly suppressed by interference
effects, and has a comparatively unfavorable polar angle
distribution. We also found that it is
necessary to employ a different cut on
the lepton energies than would be used in the neutralino LSP scenario
to separate the signal from the $WW$ backgrounds,
because of the kinematic characteristics of the decay $\slepton\rightarrow
\ell\G$ with a nearly massless gravitino. The pair production of
$\NI\NI$ can also lead to spectacular signatures involving two energetic
leptons with the same charge (and two softer leptons with the opposite
charge), in both the slepton co-NLSP and stau NLSP scenarios. Long slepton
lifetimes should lead to a relatively easy discovery from observation of
heavy charged particle tracks and/or decay kinks. One of the more difficult
scenarios involves a stau which is much lighter than all of the other
superpartners and which decays promptly into $\tau\G$. In this case
the only discovery signal is $\tau^+\tau^-\Etot$ with a significant background
from $WW$ production.

It should be noted that the signals we have studied are considerably more
general than the models outlined in section II.
The same scenarios and qualitative features of the
signals arise in a much larger class of GMSB models
with, for example, different numbers of messenger fields
and/or widely different messenger scales.
In most cases the discovery of a GMSB signal will be readily
distinguishable
from the predictions of models with a neutralino LSP.
Therefore the variety of different signal possibilities points
to the exciting prospect of
simultaneously discovering supersymmetry  and uncovering some of the most
prominent features of the
mechanism of supersymmetry breaking.


Acknowledgments: We are grateful to D.~Casta\~no, R.~Faccini,
G.~Kane, G.~Mahlon, and D.~Stuart for helpful discussions.
This work was supported in part by the U.S. Department of Energy.
The work of S.A. was supported mainly by an INFN postdoctoral fellowship,
Italy.


\newpage

\begin{table}
\renewcommand{\arraystretch}{1.2}\small\normalsize
\begin{center}
\begin{tabular}{c|rrr} \hline\hline
\multicolumn{4}{r}{Cross section (fb) for
$\sum\limits_{i=e,\mu,\tau} \ph\ph\nu_i\overline\nu_i$
after} \\ \cline{1-1}
$\sqrt{s}$ (GeV) & cuts: \ \ (1--2) & (1--3) & (1--4) \\ \hline
172 & 131 & 65.6 & 0.79 \\
190 & 102 & 70.0 & 0.80 \\
\hline\hline
\end{tabular}
\end{center}
\caption{The $\protect\ph\protect\ph\nu\bar{\nu}$
background to the $\gamma\gamma\E$ signal at LEP172 and LEP190,
after the following cuts (described in the text):
(1) $|\cos \theta_{\protect\ph}| < 0.95$,
(2) $(p_T)_{\protect\ph} > 0.0325 \protect\sqrt{s}$,
(3) $E_{\protect\ph} > {\textstyle\protect\frac{1}{4}}
( \protect\sqrt{s} - \protect\sqrt{s - (140 \; {\rm GeV})^2} )$,
(4) $M_{\rm INV} < 80 \; {\rm GeV}$.
The missing invariant mass cut (4) clearly reduces the main background
to a negligible
level. The additional cut $M_{\rm INV} > 5$ GeV is needed to eliminate the
background from $\gamma\gamma(\gamma)$, where
$(\gamma)$ is lost in either the detector or the beam pipe.}
\label{phph-table}
\end{table}

\begin{table}
\renewcommand{\arraystretch}{1.2}\small\normalsize
\begin{center}
\begin{tabular}{l|rrrr} \hline\hline
& \multicolumn{4}{c}{Cross section (fb) at LEP190 after} \\
Background ($\ell = e$ or $\mu$) & cuts: \ \
(1--3) & (1--4) & (1--5) & (1--6)
  \\ \hline
a) $W^+W^- \ra \ell^+\ell^- \nu_\ell \overline{\nu}_\ell$
& 163  & 141  & 140  &  22.3 \\
b) $W^+W^- \ra \ell^\pm \tau^\mp (\ra \ell^\mp \nu_\tau \nu_\ell)
  \nu_\ell \nu_\tau$
&  57.2 &  42.5   &  17.7 & 3.03 \\
c) $\sum\limits_{i=e,\mu,\tau} \gamma^* (\ra \ell^+ \ell^-)
                                   Z (\ra \nu_i \overline{\nu}_i)$
& 31.2 & 2.97 & 0.83  & 0.40 \\
d) $\sum\limits_{i=e,\mu,\tau} e^+e^- Z (\ra \nu_i \overline{\nu}_i)$
&      &       &   &  \\
$\phantom{d)}$ {\small (other than $ZZ$, $\gamma^*Z$ contribs.)}
& 14.8 &  0.93 & 0.56 & 0.13 \\
e) $\sum_{\pm} e^\pm \nu_e W^\mp (\ra e^\mp \nu_e)$
& 13.9 & 11.4  & 7.69 & 1.96 \\
f) $\sum\limits_{i=e,\mu,\tau} Z (\ra \ell^+ \ell^-)
Z (\ra \nu_i \overline{\nu}_i)$

& 4.92 &  0.27 & 0.27 &  0.13 \\
\hline\hline
\end{tabular}
\end{center}
\caption{The dominant backgrounds to dilepton signals
at $\protect\sqrt{s} = $ 190 GeV
after the following cuts (described in the text):
(1) $|\eta_\ell| < 2.5$,
(2) $\protect\slashchar{p}_T > 0.05 \protect\sqrt{s}$,
(3) $\cos \phi(\ell^+\ell^-) > -0.9$,
(4) $|M_{\rm INV} - M_Z| > 10$~GeV,
(5) $E_\ell > 20$~GeV,
(6) $\pm \cos \theta_{\ell^\pm} > 0$.
Other channels [e.g. $\nu\overline\nu Z(\ra \ell^+\ell^-)$;
$\mu^\pm \nu_\mu W^\mp (\ra \mu^\mp \nu_\mu)$; contributions to
$\mu^+\mu^-Z$ from
processes other than $\gamma^*Z$ and $ZZ$ production; multiperipheral
diagrams; etc.]
produce additional backgrounds at the level of 1 fb or less, before cuts.
Cuts (1-2) are needed to reduce backgrounds from $e^+e^-(\gamma)$,
$e^+e^-(\ell^+\ell^-)$, where $(x)$ means $x$ is lost in the beam pipe,
as well as to ensure final state detection.  Cut (3) is necessary to
eliminate the background from $\tau^+(\ra \ell^+ \nu \nu)
\tau^- (\ra \ell^- \nu \nu)$ and is also quite effective in reducing
a), b), d) and e) [by about 24\%, 23\%, 18\% and 15\% of the
corresponding respective amounts after cuts (1-2)]. Note that the
processes d) and e) are backgrounds only for the $e^+e^-\E$ signal.}
\label{lepton-table}
\end{table}

\begin{table}
\renewcommand{\arraystretch}{1.4}\small\normalsize
\begin{small}
\begin{center}
\begin{tabular}{ccll} \hline\hline
         & Sparticle  & & \\
Scenario & Production & ~~~~Signal & Comments \\ \hline
neutralino NLSP  & $\NI\NI$
 & $\left\{ \begin{array}{l}
            \gamma\gamma\E \\
            \mbox{displaced} \; \gamma\mbox{s}
            \end{array}\right.$
 & $\NI \ra \gamma\G$ decays are
   $\left\{ \begin{array}{l}
            \mbox{prompt} \\
            \mbox{within the detector}
            \end{array}\right.$ \vspace*{1.0mm} \\ \cline{3-4}
 & $\lRp\lRm$ & $\left\{ \begin{array}{l} \gamma\gamma\ell^+\ell^-   \E \\
                 \ell^+\ell^- + \mbox{displaced} \; \gamma\mbox{s}
                 \end{array} \right. $
 & (as above) \vspace*{1.0mm} \\ \hline
slepton co-NLSP
 & $
    \lRp\lRm 
    $
 & $\left\{ \begin{array}{l} \ell^+\ell^-   \E \ \ ^{(*)} \\
            \tilde{\ell} \ra \ell\G \; \mbox{decay kinks} \\
            \mbox{charged} \; \tilde{\ell} \; \mbox{tracks} \;
            \end{array}\right.$
 & $\tilde{\ell} \ra  \ell\G$ decays are
   $\left\{ \begin{array}{l} \mbox{prompt} \\
            \mbox{within the detector} \\
            \mbox{outside the detector} \end{array}\right.$
   \vspace*{1.0mm} \\ \cline{3-4}
 & $\NI\NI$
 & $\left\{ \begin{array}{l}
            \ell^+ {\ell'}^+ ( \ell^- {\ell'}^- )\E, \vspace*{-2.0mm}\\
       \ell^- {\ell'}^- ( \ell^+ {\ell'}^+ )\E,\>\mbox{and}\vspace*{-2.0mm}\\
                    \ell^+ {\ell'}^- ( \ell^- {\ell'}^+ )   \E
            \end{array}\right.$
 & $(\ell \ell')$ leptons are soft or undetected \vspace*{1.0mm} \\
 \hline
stau NLSP & $\tilde{\tau}_1^+\tilde{\tau}_1^-$
 & $\left\{ \begin{array}{l} \tau^+\tau^-   \E \\
            \tilde{\tau}_1 \ra \tau\G \; \mbox{decay kinks} \\
            \mbox{charged} \; \tilde{\tau}_1 \; \mbox{tracks} \;
            \end{array}\right.$
 & $\tilde{\tau}_1 \ra \tau\G$ decays are
   $\left\{ \begin{array}{l} \mbox{prompt} \\
            \mbox{within the detector} \\
            \mbox{outside the detector} \end{array}\right.$
   \vspace*{1.0mm} \\ \cline{3-4}
 & $\NI\NI$
 & $\left\{ \begin{array}{l}
            \tau^+ {\tau}^+ ( \tau^- {\tau}^- ) \E, \vspace*{-2.0mm}\\
          \tau^- {\tau}^- ( \tau^+ {\tau}^+ ) \E,\>\mbox{and}\vspace*{-2.0mm}\\
          \tau^+ {\tau}^- ( \tau^+ {\tau}^- )   \E
            \end{array}\right.$
 & $\begin{array}{l}(\tau \tau) \ \ 
                 \mbox{leptons are soft or undetected;}\\
                 \mbox{possibly with 2 or 4 additional soft}\\
                  \mbox{leptons}\ (e\ \mbox{or}\ \mu)\ 
                  \mbox{if}\ \ 
                  \NI\rightarrow\ell\slepton_R\ \  \mbox{is open} \end{array}$
   \vspace*{1.0mm} \\ \cline{3-4}
 & $\lRp\lRm$
 & $\tau^+\tau^- (\ell^+\ell^- \tau^+\tau^-)   \E$
 & $(\ell^+ \ell^- \tau^+ \tau^-)$ leptons are
   soft or undetected \vspace*{1.0mm} \\ \hline
$\begin{array}{c} \mbox{neutralino-stau} \\ \mbox{co-NLSP} \end{array}$
 & $\begin{array}{l}
    \NI\NI \> , \\
    \tilde{\tau}_1^+\tilde{\tau}_1^-
    \end{array}$
 & $\left\{ \begin{array}{l}
   \gamma\gamma   \E,\>\> \tau^+\tau^-   \E \\
   \mbox{displaced} \; \gamma\mbox{s and} \vspace*{-2.0mm} \\
   \tilde{\tau}_1 \ra \tau\G \; \mbox{decay kinks} \\
   \mbox{charged} \; \tilde{\tau}_1 \; \mbox{tracks}
   \end{array}\right.$
 & NLSP decays are $\left\{ \begin{array}{l}
                      \mbox{prompt} \\
                      \mbox{within the detector} \\
                      \mbox{outside the detector}
                      \end{array}\right.$ \vspace*{1.0mm} \\ \hline\hline
\end{tabular}
\end{center}
\end{small}
\caption{The possible signatures at LEP2 in the four different
NLSP scenarios.  
The notation (1) ``prompt'', (2) ``within the detector'', and (3)
``outside the detector'' refers to a NLSP decay such that the
decay vertex is (1) close to the interaction region
and not measurably displaced, (2) possibly resolvable
with a detector, and (3) well outside the detector. In the
neutralino NLSP and slepton co-NLSP cases, $\ell$ stands for
$e$, $\mu$, or $\tau$; 
in the stau NLSP case $\ell$ stands for $e$ or $\mu$.
$^{(*)}$ In the slepton
co-NLSP case, the particular $\ell^+\ell^-\E$ signature which is
most likely to be observable is $\mu^+\mu^-\E$, as explained in 
Section III.B.}
\label{sig-table}
\end{table}
\end{document}